\documentclass[10pt, DIV=14]{scrartcl}

\usepackage{microtype}
\usepackage{amsmath}
\usepackage{amssymb}
\usepackage{hyperref}
\usepackage{graphicx}
\usepackage{xcolor}
\usepackage{booktabs}
\usepackage{multirow}
\usepackage{makecell}
\usepackage{amsthm}
\usepackage{enumitem}
\usepackage{setspace}
\usepackage{authblk}

\definecolor{Pastel1-1}{HTML}{fbb4ae}
\definecolor{Pastel1-2}{HTML}{b3cde3}

\definecolor{Blues3-1}{HTML}{deebf7}
\definecolor{Blues3-2}{HTML}{9ecae1}
\definecolor{Blues3-3}{HTML}{3182bd}

\usepackage[labelfont={sf, bf}, font=small, format=plain]{caption}

\usepackage{mathrsfs}

\usepackage[biblabel]{cite}

\usepackage[labelsep=period]{caption}

\author[1]{Fabian Fritz\thanks{Corresponding author.}\textsuperscript{,}}
\author[2]{Stefan Adami}
\author[1, 2]{Nikolaus A. Adams}

\affil[1]{Chair of Aerodynamics and Fluid Mechanics, Technical University of Munich (TUM), Boltzmannstr. 15, Garching, 85748, Germany}
\affil[2]{Munich Institute of Integrated Materials, Energy and Process Engineering (MEP), Technical University of Munich (TUM), Lichtenbergstr. 4a, Garching, 85748, Germany}

\title{A conservative Godunov-type mesh-free hydrodynamics scheme for weakly compressible multiphase flows\thanks{Submitted to \emph{Journal of Computational Physics}.}}

\usepackage{draftwatermark}
\SetWatermarkText{\normalfont\textsf{PREPRINT}}
\SetWatermarkScale{5}
\SetWatermarkColor[gray]{0.95}

\begin{document}

\maketitle
{\renewcommand{\thefootnote}{}\footnotetext{\textit{E-mail address:} fabian.fritz@tum.de (F.\ Fritz)}
\begin{abstract}
    We present a Godunov-type mesh-free hydrodynamics scheme for the numerical simulation of weakly compressible single- and multiphase flows.
The method employs a consistent second-order spatial reconstruction of the primitive variables at the inter-particle interface, applies slope limiting, and solves the moving Riemann problem using a subsonic HLLC approximate Riemann solver.
Particles are advected with a quasi-Lagrangian material velocity incorporating an inherent self-relaxation mechanism that suppresses the formation of anisotropic particle distributions.
Near material interfaces, one-sided renormalized gradients are used to handle density discontinuities and slope limiting is applied for each phase independently.
The Lagrangian character of material interfaces is preserved by constraining the relaxation mechanism to act tangentially to the interface.
For same-phase particle interactions, the Riemann problem is centered at the arithmetic mean of the two material velocities.
For cross-phase interactions, it is centered at the contact-wave velocity.
This choice naturally eliminates spurious mass fluxes across phase boundaries.
The methodology is validated on a set of challenging benchmarks, including the inviscid Taylor-Green vortex, material interface advection, surface gravity waves near hydrostatic equilibrium, the Kelvin-Helmholtz instability, and the Rayleigh-Taylor instability under gravitational stratification.

\end{abstract}
\section{Introduction}
Multiphase flows are of prime interest in many engineering applications and natural phenomena, such as, e.g., the detrimental effect of hydrodynamic instabilities in inertial confinement fusion~\cite{betti2016inertial, zhou2025instabilities, atzeni2005fluid}, the injection and atomization of a liquid fuel jet in supersonic crossflow~\cite{kuhn2022experimentally, kuhn2021all}, laser-induced cavitation for surface cleaning using microjetting~\cite{ohl2006surface, bussmann2023investigation}, and the rapid evaporation of liquid metal in laser-based powder-bed fusion~\cite{bidare2018fluid}.
Their importance, along with the continuous growth in computational power, has driven the development of advanced numerical methods for computational fluid dynamics (CFD) to complement experimental research.
\par
A sharp interface between phases is a moving discontinuity.
The numerical representation and evolution of a moving discontinuity on a fixed mesh can exhibit artifacts due to numerical dissipation, see, e.g.,~\cite{saurel1999godunov,so2012anti}, motivating numerical methods in which the discretization follows the fluid motion.
Meshfree (or meshless) methods are a class of numerical methods in computational mechanics, including fluid, solid, and structural mechanics.
They represent the computational domain using nodes or particles without a predefined mesh topology.
This approach is advantageous for problems involving large deformations, moving interfaces, and evolving geometries, see, e.g.,~\cite{belytschko1996meshless, li2002meshfree}.
In CFD, such methods are categorized as Lagrangian particle methods, in contrast to Eulerian grid-based methods, see, e.g., Agertz et al.~\cite{agertz2007fundamental}.
However, recent developments in CFD have given rise to Eulerian formulations of particle methods (e.g., Eulerian SPH~\cite{lind2016high,nasar2019eulerian,nasar2021high}) and moving-mesh schemes on unstructured, dynamically evolving Voronoi grids, including finite-volume (FV)~\cite{springel2010galilean,weinberger2023modelling} and discontinuous Galerkin (DG)~\cite{mocz2014discontinuous,gaburro2020high} variants.
We therefore adopt the broader and more pragmatic distinction between mesh-free hydrodynamics (MFH) and grid-based hydrodynamics\footnote{Strictly speaking, \emph{hydrodynamics} refers to the dynamics of water-like fluids, but the term has been adopted in the astrophysics community to denote \emph{fluid dynamics}.
We follow this convention and use \emph{hydrodynamics} interchangeably with \emph{fluid dynamics}.}.
MFH schemes avoid discretizing the computational space into a mesh.
Instead, the fluid is represented by points (also referred to as particles or material points) that carry properties such as mass, momentum, and energy.
\par
Smoothed particle hydrodynamics (SPH)~\cite{lucy1977numerical, gingold1977smoothed} is the most widely used MFH scheme due to its algorithmic simplicity and favorable properties.
SPH uses a smoothing kernel to approximate a function and its spatial gradient from interactions between neighboring particles.
SPH is a fully Lagrangian method, i.e., the particles move with the local fluid velocity or an approximation of it.
It is Galilean invariant and conserves mass, linear momentum, angular momentum, energy, and entropy in its original form~\cite{dehnen2012improving}.
However, standard SPH is not zeroth-order consistent (referred to as ``E0-error'')~\cite{morris1996analysis, dehnen2012improving, litvinov2015towards, read2010resolving, zhang2025towards}, lacks sufficient intrinsic numerical dissipation~\cite{monaghan1983shock, vila1999particle, monaghan1992smoothed}, does not converge for a fixed number of neighboring particles~\cite{springel2010smoothed, zhu2015numerical, quinlan2006truncation}, and suffers from numerical instabilities such as tensile instabilities\footnote{The tensile instability occurs when particles experience negative pressure (or tension) and spurious attractive forces lead to particle clumping and particle voids.}~\cite{monaghan2000sph,swegle1995smoothed} and pairing instabilities\footnote{The pairing (or clumping) instability is a numerical artifact, where particles form close pairs, reducing the effective resolution.}~\cite{dehnen2012improving}.
It also produces anisotropic particle distributions~\cite{oger2016sph, adami2013transport} that reduce the accuracy and lead to stability issues.
These shortcomings have motivated a large body of work aimed at improving SPH.
Here, we consider SPH methods that employ kernel-gradient estimates to approximate spatial derivatives, see, e.g.,~\cite{hu2006multi, zhang2017weakly, adami2010new, adami2013transport, adami2012generalized}.
Hopkins~\cite{hopkins2015new} and Gaburov and Nitadori~\cite{gaburov2011astrophysical} proposed a new class of Godunov-type mesh-free hydrodynamics (GMFH) schemes to address the accuracy and stability deficiencies of SPH for applications in astrophysics.
The methodology is based on the theoretical work of Lanson and Vila~\cite{lanson2008renormalized,lanson2008renormalized2} and introduces renormalized moving-least-squares gradient estimates that are second-order accurate to improve on consistency and convergence issues in classical SPH formulations.
A conservative flux formulation is employed, similar to that of moving FV schemes~\cite{springel2010galilean}.
This allows for the natural inclusion of an intrinsic numerical dissipation mechanism through the introduction of an appropriate Riemann problem.
As pointed out by Hopkins~\cite{hopkins2015new}, the GMFH scheme encompasses a true arbitrary Lagrangian-Eulerian (ALE) scheme, referred to as the meshless finite-volume (MFV) method, and a genuinely Lagrangian scheme, referred to as the meshless finite-mass (MFM) method.
The methodology of Lanson and Vila~\cite{lanson2008renormalized,lanson2008renormalized2} is not the only way to improve on the accuracy of SPH.
Conservative reproducing kernel SPH (CRKSPH) of Frontiere et al.~\cite{frontiere2017crksph} is a closely related reformulation that achieves zeroth- and first-order consistency while remaining conservative for mass, momentum, and energy.
A CRKSPH gradient estimator has since been integrated into the open-source code GIZMO~\cite{hopkins2015new, springel2002cosmological}, unifying the MFM and MFV framework with CRKSPH.
\par
A central challenge in MFH schemes is the control of numerical dissipation.
Lagrangian SPH treats advection exactly, reducing the need for numerical dissipation to ensure convective numerical stability.
The absence of upwinding means strong discontinuities, such as shocks and contact discontinuities, and acoustic waves are not stable when physical viscosity vanishes.
Monaghan and Gingold~\cite{monaghan1983shock} addressed this by adding an artificial viscosity term to the momentum equation.
This leads to excessive dissipation of smooth-flow features.
Density-diffusion schemes~\cite{molteni2009simple, antuono2010free} shift the explicit dissipation term into the continuity equation, confining its effect to the density and pressure fields.
Nonetheless, the approach still relies on explicit numerical viscosity.
In GMFH schemes, numerical dissipation is introduced implicitly by solving a Riemann problem at each inter-particle interface, following the pioneering work of Vila~\cite{vila1999particle} and Inutsuka~\cite{inutsuka2002reformulation}.
To minimize dissipation of smooth-flow features, a high-order MUSCL-type reconstruction~\cite{vanleer1979towards, toro2009riemann} of the fluid state at the interface is typically used.
This adds significant computational overhead and has motivated efforts to recover accuracy without it.
Ferrari et al.~\cite{ferrari2009free} proposed combining a monotone upwind flux for the continuity equation with a non-diffusive central flux for the momentum equation.
Zhang et al.~\cite{zhang2017weakly} introduced a dissipation limiter that heuristically reduces intrinsic numerical dissipation, recovering accuracy for weakly compressible flows at lower computational cost.
\par
A second challenge, closely coupled to dissipation control, is the tendency of Lagrangian MFH schemes to produce anisotropic particle distributions, especially for weakly compressible flows.
These irregular structures decrease the accuracy~\cite{litvinov2015towards, zhang2025towards} and can cause spurious numerical instabilities if the smoothing length does not adapt to the particle number density.
To mitigate this, many current schemes use a different material velocity (also referred to as transport velocity) than the fluid velocity. Two principal approaches exist:
In the first approach, Monaghan~\cite{monaghan1989problem} proposed to compute the material velocity by smoothing the fluid velocity to prevent particle clustering~\cite{monaghan1992smoothed, monaghan2002sph}.
This approach is Galilean invariant and conserves angular momentum~\cite{michel2022particle}, but it does not prevent anisotropic particle distributions.
In the second approach, the material velocity equals a regularized fluid velocity, which introduces a small perturbation that moves particles slightly from their Lagrangian trajectory into regions of lower particle number densities~\cite{lind2012incompressible, xu2009accuracy, adami2013transport}.
This approach significantly increases the accuracy and stability~\cite{litvinov2015towards} of weakly compressible MFH schemes and has been continuously refined, e.g., for free-surface flows~\cite{sun2019consistent, zhang2017generalized} and multiphase flows~\cite{rezavand2020weakly, mokos2016multiphase}.
Incorporating a quasi-Lagrangian material velocity with an inherent particle relaxation mechanism naturally fits the ALE property of contemporary MFH schemes, as demonstrated by Oger et al.~\cite{oger2016sph}.
\par
Extending these advances to multiphase flows introduces additional modeling challenges.
Colagrossi and Landrini~\cite{colagrossi2003numerical} proposed one of the first multiphase SPH formulations capable of handling large density contrasts.
However, their approach does not conserve mass.
Hu and Adams~\cite{hu2006multi} addressed this shortcoming by introducing a density formulation, where neighboring particles contribute only to the specific volume, a quantity continuous across phase interfaces, rather than to the density directly.
This accommodates density discontinuities at material interfaces while preserving mass conservation.
The framework was later extended to incompressible multiphase flows~\cite{hu2009constant, hu2007incompressible}.
Grenier et al.~\cite{grenier2009hamiltonian} developed a closely related multiphase SPH method grounded on a Lagrangian variational principle, following Bonet and Lok~\cite{bonet1999variational}.
This variational formulation guarantees that the resulting particle system is Hamiltonian, with linear and angular momentum conserved by construction, and features a thermodynamically consistent pressure-gradient formulation.
Grenier et al.~\cite{grenier2009hamiltonian} also provide a formal justification for the pressure gradient used by Hu and Adams~\cite{hu2006multi}.
The formulation further generalizes the method to free-surface flows, where kernel summation for the specific volume is ill-posed near the free surface.
This is resolved by using a renormalized kernel and evolving the particle volume via a continuity equation.
\par
The foundational work of Colagrossi and Landrini~\cite{colagrossi2003numerical}, Hu and Adams~\cite{hu2006multi}, and Grenier et al.~\cite{grenier2009hamiltonian} paved the way for increasingly capable multiphase models. 
Monaghan and Rafiee~\cite{monaghan2013simple} later proposed a simplified multiphase SPH algorithm to reduce the complexity of earlier formulations while retaining their essential properties.
Rezavand et al.~\cite{rezavand2020weakly} proposed a quasi-Lagrangian multiphase SPH method that is based on the low-dissipation Riemann solver of Zhang et al.~\cite{zhang2017weakly} and the transport velocity formulation of Adami et al.~\cite{adami2013transport}.
\par
The consistent treatment of weakly compressible multiphase flows within a GMFH scheme remains an open problem.
At its core lies a fundamental contradiction, as conserving individual particle masses requires a purely Lagrangian advection velocity, but purely Lagrangian schemes tend to produce anisotropic particle distributions that reduce accuracy and stability.
Relaxing the Lagrangian constraint introduces a non-Lagrangian advection velocity and mass fluxes between material points.
When these fluxes cross a material interface with a large density gradient, they may cause severe numerical instabilities.
Therefore, a quasi-Lagrangian advection velocity that corrects anisotropic particle distributions while avoiding numerical instabilities due to  mass fluxes across the material interfaces is essential for a robust multiphase GMFH scheme.
In current state-of-the-art weakly compressible multiphase SPH methods, e.g.,~\cite{rezavand2020weakly, mokos2016multiphase}, this challenge is addressed by keeping individual particle masses constant, while also restricting particle relaxation to the light phase.
Yet these choices both present issues.
One is that fixing particle masses within a quasi-Lagrangian advection scheme is formally inconsistent.
Another is that restricting relaxation to the light phase is largely heuristic, as its validity across different density ratios remains unclear, and it fails to prevent the formation of anisotropic particle structures in the heavy phase.
\par
In this work, we propose a mesh-free method for the numerical simulation of weakly compressible single- and multiphase flows that addresses this fundamental problem.
In the following, we summarize our contribution and outline the work.
\begin{itemize}
    \item %
        We derive a unified GMFH scheme for hyperbolic conservation laws following the theoretical work of Ivanova et al.~\cite{ivanova2013common}.
        We show that the scheme generalizes a subset of existing MFH schemes, including the multiphase SPH formulation of Hu and Adams~\cite{hu2006multi} and the renormalized gradient formulation of Gaburov and Nitadori~\cite{gaburov2011astrophysical}.
    \item %
        We introduce a single-phase GMFH scheme for weakly compressible Euler equations following the work of Lanson and Vila~\cite{lanson2008renormalized, lanson2008renormalized2}, Gaburov and Nitadori~\cite{gaburov2011astrophysical}, and Hopkins~\cite{hopkins2015new,hopkins2015accurate,hopkins2017anisotropic} for compressible hydrodynamics.
        We apply a second-order spatial reconstruction of the primitive variables at the inter-particle interface, where a multidimensional limiter is utilized to assure that the reconstruction is monotonic and non-oscillatory.
        The moving Riemann problem at the inter-particle interface is solved using the subsonic HLLC approximate Riemann solver.
        A second-order least-squares matrix gradient operator is used to approximate the inter-particle interface area and the gradient of the primitive variables.
        We use a quasi-Lagrangian material velocity to enhance stability and accuracy for weakly compressible flows by improving the particle distribution through an intrinsic self-relaxation mechanism~\cite{michel2022particle, litvinov2015towards}.
        The discrete transport equations are integrated in time using a drift-kick-drift (DKD) leapfrog integration scheme with a Runge-Kutta (RK) kick integrator for the conservative states.
    \item %
        We extend the single-phase GMFH scheme to multiphase flows.
        In the vicinity of the material interface, we adopt a one-sided renormalized gradient formulation to account for the density discontinuity, and restrict the slope limiting to the respective phase.
        The material velocity is adjusted so that the relaxation mechanism acts only in the tangential direction of the interface in its vicinity, thereby preserving the natural Lagrangian interface motion.
        The inter-particle interface velocity of the Riemann problem is the arithmetic mean of the bulk material velocities away from the interface and the contact velocity across it, ensuring that no spurious mass fluxes cross the material interface.
    \item %
        We investigate the proposed methodology using challenging single- and multiphase problems, including the inviscid Taylor-Green vortex, a material interface advection problem, the surface gravity wave problem near hydrostatic equilibrium, the Kelvin-Helmholtz instability, and, finally, a Rayleigh-Taylor instability under gravitational stratification.
\end{itemize}
\section{Mathematical model}\label{sec:MathematicalModel}
The fluid motion of an inviscid weakly compressible flow is described through the Euler equations, which, expressed as a balance law, read
\begin{equation}\label{eq:ConservationLaw}
    \tfrac{\partial}{\partial t}(\mathbf{Q}) + \mathrm{div}(\boldsymbol{\mathsf{F}}(\mathbf{Q})) = \mathbf{S}(\mathbf{Q}).
\end{equation}
The vector of conserved densities $\mathbf{Q}$, the tensorial hyperbolic flux function $\boldsymbol{\mathsf{F}}$, and the source vector $\mathbf{S}$, are
\begin{equation}
    \mathbf{Q}
    =
    \begin{pmatrix}
        \rho \\
        \rho\mathbf{v}
    \end{pmatrix},
    \quad
    \boldsymbol{\mathsf{F}}(\mathbf{Q})
    =
    \begin{pmatrix}
        \rho\mathbf{v} \\
        \rho\mathbf{v}\otimes\mathbf{v} + p\mathbf{I}
    \end{pmatrix},
    \quad
    \mathbf{S}(\mathbf{Q})
    =
    \begin{pmatrix}
        0 \\
        \rho\mathbf{f}
    \end{pmatrix},
\end{equation}
where $\rho$ is the mass density, $\rho\mathbf{v}$ is the momentum density, $\mathbf{v}$ is the fluid velocity, $p = p(\rho)$ is the pressure and $\mathbf{f}$ is an external force density (or acceleration) field, e.g., gravitational acceleration.
The vector of primitive variables is
\par
\begin{equation}\label{eq:PrimitiveVariables}
    \mathbf{W}
    =
    \begin{pmatrix}
        \rho \\
        \mathbf{v}
    \end{pmatrix}.
\end{equation}
The pressure and density are coupled by the weakly compressible equation-of-state (EOS)
\begin{equation}\label{eq:EquationOfState}
    p(\rho) = c_{0}^{2}(\rho - \rho_{0}) + p_{b},
\end{equation}
where $c_{0}$ is an artificial speed of sound, $\rho_{0}$ is the reference density and $p_b$ is a background pressure.
\par
The balance law (Eq.~\eqref{eq:ConservationLaw}) describes the conservation of total mass $M$ and total linear momentum $\mathbf{P}$ in the absence of sources, i.e., $\mathbf{f} = \mathbf{0}$, such that
\begin{equation}
    \tfrac{\mathrm{d}}{\mathrm{d}t}
    \begin{pmatrix}
        M \\ \mathbf{P}
    \end{pmatrix}
    =
    \tfrac{\mathrm{d}}{\mathrm{d}t}\left(\int \mathbf{Q}\,\mathrm{d}\mathbf{r}\right) = 0.
\end{equation}
\section{Numerical method}\label{sec:NumericalMethod}
We introduce the weighting function $\chi_{i}(\mathbf{r})$ at the material coordinates $\mathbf{r}_{i}$, $i = 1,\hdots,N$, where $N$ is the number of material points (or particles).
The weighting function satisfies a partition of unity property
\begin{equation}\label{eq:PartitionOfUnity}
    \sum_{i} \chi_{i}(\mathbf{r})
    =
    1,
\end{equation}
and consequently
\begin{equation}\label{eq:PartitionOfNullity}
    \sum_{i} \nabla\chi_{i}(\mathbf{r})
    =
    0.
\end{equation}
The effective volume is defined as
\begin{equation}\label{eq:EffectiveVolume}
    V_{i}(t)
    \equiv
    \int \chi_{i}(\mathbf{r})\,\mathrm{d}\mathbf{r},
\end{equation}
such that $V = \sum_i V_i(t)$, where $V$ is the total volume.
The average of the vector of conserved densities, denoted by $\overline{\mathbf{Q}}_{i}(t)$, over the effective volume (Eq.~\eqref{eq:EffectiveVolume}) is defined, such that
\begin{equation}\label{eq:AverageConservatives}
    \overline{\mathbf{Q}}_{i}(t) \equiv \frac{1}{V_{i}(t)}\int \mathbf{Q}(\mathbf{r}, t)\chi_{i}(\mathbf{r})\,\mathrm{d}\mathbf{r},
\end{equation}
following Hu and Adams~\cite{hu2006multi}, and Serrano and Espanol~\cite{serrano2000thermodynamically}.
The time derivative of Eq.~\eqref{eq:AverageConservatives} is given by means of the product rule through
\begin{equation}\label{eq:TimeDerivativeOfConservatives}
    \tfrac{\mathrm{d}}{\mathrm{d}t}(\overline{\mathbf{Q}}_{i})(t)
    =
    -\frac{\dot{V}_{i}}{V_{i}}\overline{\mathbf{Q}}_{i}(t)
    +\frac{1}{V_{i}}\int\tfrac{\partial}{\partial t}(\mathbf{Q})(\mathbf{r}, t)\chi_{i}(\mathbf{r})\,\mathrm{d}\mathbf{r}
    +\frac{1}{V_{i}}\int\mathbf{Q}(\mathbf{r}, t) \tfrac{\partial}{\partial t}(\chi_{i})(\mathbf{r})\,\mathrm{d}\mathbf{r}.
\end{equation}
We can use the product rule on the first term of the right-hand side (RHS) and the balance law (Eq.~\eqref{eq:ConservationLaw}) on the second term of the RHS to reformulate Eq.~\eqref{eq:TimeDerivativeOfConservatives} into a transport equation.
The transport equation describes the conservation of the extensive variables, e.g., mass, momentum or energy, such that
\begin{equation}
    \tfrac{\mathrm{d}}{\mathrm{d}t}(\overline{\mathbf{Q}}_{i} V_{i})(t)
    =
    -\underbrace{\int\mathrm{div}\,\boldsymbol{\mathsf{F}}(\mathbf{Q})\chi_{i}(\mathbf{r})\,\mathrm{d}\mathbf{r}}_{(\spadesuit)}
    +\underbrace{\int\mathbf{Q}(\mathbf{r}, t)\tfrac{\partial}{\partial t}({\chi}_{i})(\mathbf{r})\,\mathrm{d}\mathbf{r}}_{(\bigstar)}
    +\underbrace{\int \mathbf{S}(\mathbf{Q})\chi_{i}(\mathbf{r})\,\mathrm{d}\mathbf{r}}_{(\clubsuit)},
\end{equation}
which contains three contributions to the RHS:
The first integral $(\spadesuit)$ describes the change of the extensive conservative variables due to the flux exchange between adjacent volumes.
We utilize Eq.~\eqref{eq:PartitionOfUnity} and Eq.~\eqref{eq:PartitionOfNullity} to derive the identity
\begin{subequations}\label{eq:GradIdentity}
\begin{align}
    \nabla\chi_{i}(\mathbf{r})
    &=
    \nabla\chi_{i}(\mathbf{r})\sum_{j}\chi_{j}(\mathbf{r}) - \chi_{i}(\mathbf{r})\sum_{j}\nabla\chi_{j}(\mathbf{r})
    \\
    &=
    \sum_{j} \big(\chi_{j}(\mathbf{r})\nabla\chi_{i}(\mathbf{r}) - \chi_{i}(\mathbf{r})\nabla\chi_{j}(\mathbf{r})\big)
\end{align}
\end{subequations}
to show that
\begin{subequations}
\begin{align}
    -\int \mathrm{div}\,\boldsymbol{\mathsf{F}}(\mathbf{Q})\,\chi_{i}(\mathbf{r})\,\mathrm{d}\mathbf{r}
    &=
    \int \boldsymbol{\mathsf{F}}(\mathbf{Q})\bullet\nabla\chi_{i}(\mathbf{r})\,\mathrm{d}\mathbf{r}\phantom{\sum_{j}} \\
    &=
    \sum_{j}\int \boldsymbol{\mathsf{F}}(\mathbf{Q})\bullet\big(\chi_{j}(\mathbf{r})\nabla\chi_{i}(\mathbf{r})-\chi_{i}(\mathbf{r})\nabla\chi_{j}(\mathbf{r})\big)\,\mathrm{d}\mathbf{r} \\
    &\equiv
    -\sum_{j} \int \boldsymbol{\mathsf{F}}(\mathbf{Q})\bullet\mathrm{d}\boldsymbol{\Sigma}_{ij}.
\end{align}
\end{subequations}
We introduce a second-order approximation to the last integral using a one-point quadrature and replace the physical flux function $\boldsymbol{\mathsf{F}}(\mathbf{Q})$ by a numerical flux function $\overline{\boldsymbol{\mathsf{F}}}_{ij} = \overline{\boldsymbol{\mathsf{F}}}(\overline{\mathbf{Q}}_{i}, \overline{\mathbf{Q}}_{j})$, such that
\begin{equation}\label{eq:OnePointQuadrature}
    \sum_{j} \int \boldsymbol{\mathsf{F}}(\mathbf{Q})\bullet\mathrm{d}\boldsymbol{\Sigma}_{ij}
    \approx
    \sum_{j} \overline{\boldsymbol{\mathsf{F}}}_{ij}\bullet\int\mathrm{d}\boldsymbol{\Sigma}_{ij}.
\end{equation}
The second integral $(\bigstar)$ describes the change of the extensive variables due to the deformation of the effective volumes from the material velocity.
The material time derivative of the weighting function is
\begin{equation}\label{eq:MaterialDerivativeOfWeightingFunction}
    \tfrac{\mathrm{d}}{\mathrm{d}t}(\chi_{i})(\mathbf{r}) = \dot{\chi}_{i}(\mathbf{r}) = \tfrac{\partial}{\partial t}(\chi_{i})(\mathbf{r}) + \dot{\mathbf{r}}_{i}(t)\bullet\nabla\chi_{i}(\mathbf{r}),
\end{equation}
where $\dot{\mathbf{r}}_{i}$ is the material velocity.
We utilize Eq.~\eqref{eq:PartitionOfUnity}, Eq.~\eqref{eq:PartitionOfNullity}, and Eq.~\eqref{eq:MaterialDerivativeOfWeightingFunction} to derive the identity
\begin{equation}\label{eq:IdentityDerivative}
    \tfrac{\partial}{\partial t}(\chi_{i})(\mathbf{r})
    =
    \sum_{j}\big(\dot{\mathbf{r}}_{j}(t)\bullet\chi_{i}(\mathbf{r})\nabla\chi_{j}(\mathbf{r})-\dot{\mathbf{r}}_{i}(t)\bullet\chi_{j}(\mathbf{r})\nabla\chi_{i}(\mathbf{r})\big)
    +
    \sum_{j}\big(\dot{\chi}_{i}(\mathbf{r})\chi_{j}(\mathbf{r}) - \dot{\chi}_{j}(\mathbf{r})\chi_{i}(\mathbf{r}) \big)
\end{equation}
such that the integral ($\bigstar$) is
\begin{align}\label{eq:BigstarIntegral}
    \int \mathbf{Q}(\mathbf{r}, t) \tfrac{\partial}{\partial t}({\chi}_{i})(\mathbf{r})\,\mathrm{d}\mathbf{r}
    &\approx
    \sum_{j} \int \mathbf{Q}(\mathbf{r}, t) \big(\dot{\mathbf{r}}_{j}(t)\bullet\chi_{i}(\mathbf{r})\nabla\chi_{j}(\mathbf{r})-\dot{\mathbf{r}}_{i}(t)\bullet\chi_{j}(\mathbf{r})\nabla\chi_{i}(\mathbf{r})\big)\,\mathrm{d}\mathbf{r}
    \\
    &\approx
    \sum_{j}\overline{\mathbf{Q}}_{ij}\otimes\overline{\dot{\mathbf{r}}}_{ij}\bullet\int\left(\chi_{i}(\mathbf{r})\nabla\chi_{j}(\mathbf{r})-\chi_{j}(\mathbf{r})\nabla\chi_{i}(\mathbf{r})\right)\,\mathrm{d}\mathbf{r}
    \\
    &\equiv
    \sum_{j}\overline{\mathbf{Q}}_{ij}\otimes\overline{\dot{\mathbf{r}}}_{ij}\bullet\int\mathrm{d}\boldsymbol{\Sigma}_{ij}.
\end{align}
where $\overline{\mathbf{Q}}_{ij}$ denotes an inter-particle average of the conserved densities, and $\overline{\dot{\mathbf{r}}}_{ij}$ denotes an inter-particle average of the material velocity.
We neglect the second sum in Eq.~\eqref{eq:IdentityDerivative}, which is antisymmetric and constitutes a rotational contribution to the interface velocity, see, e.g.,~\cite{serrano2000thermodynamically, springel2010galilean}.
The third integral $(\clubsuit)$ describes the change of the extensive conservative variables due to sources or sinks.
We approximate the term again using a second-order accurate one-point quadrature, such that
\begin{equation}
    \int \mathbf{S}(\mathbf{Q})\chi_{i}(\mathbf{r})\,\mathrm{d}\mathbf{r}
    \approx
    \mathbf{S}(\mathbf{Q}_{i}) \int \chi_{i}(\mathbf{r})\,\mathrm{d}\mathbf{r}
    =
    \mathbf{S}(\mathbf{Q}_{i}) V_{i}.
\end{equation}
The effective interface area is defined as
\begin{equation}
    \mathbf{A}_{ij}
    \equiv
    \int\mathrm{d}\boldsymbol\Sigma_{ij}
    =
    \int\big(\chi_{i}(\mathbf{r})\nabla\chi_{j}(\mathbf{r}) - \chi_{j}(\mathbf{r})\nabla\chi_{i}(\mathbf{r})\big)\,\mathrm{d}\mathbf{r}
\end{equation}
and truncated using a second-order accurate one-point quadrature, such that
\begin{subequations}\label{eq:EffectiveInterfaceArea}
\begin{align}
    \mathbf{A}_{ij}
    &=
    \int\chi_{i}(\mathbf{r})\nabla\chi_{j}(\mathbf{r})\,\mathrm{d}\mathbf{r} -\int\chi_{j}(\mathbf{r})\nabla\chi_{i}(\mathbf{r})\,\mathrm{d}\mathbf{r}
    \\
    &\approx
    \nabla\chi_{j}(\mathbf{r}_{i})\int\chi_{i}(\mathbf{r})\,\mathrm{d}\mathbf{r}
    -
    \nabla\chi_{i}(\mathbf{r}_{j})\int\chi_{j}(\mathbf{r})\,\mathrm{d}\mathbf{r}
    \\
    &=
    V_{i}\nabla\chi_{j}(\mathbf{r}_{i})
    -
    V_{j}\nabla\chi_{i}(\mathbf{r}_{j}).
\end{align}
\end{subequations}
Eventually, the transport equation for the average conservative densities is
\begin{equation}\label{eq:DiscretizedBalanceSystem}
    \tfrac{\mathrm{d}}{\mathrm{d}t}(\overline{\mathbf{Q}}_{i} V_{i})
    +
    \sum_{j} (\overline{\boldsymbol{\mathsf{F}}}_{ij} - \overline{\mathbf{Q}}_{ij}\otimes\overline{\dot{\mathbf{r}}}_{ij})\bullet\mathbf{A}_{ij}
    =
    \mathbf{S}_{i} V_{i},
\end{equation}
where $\overline{\mathbf{Q}}_{i}$ is the average of the conserved densities (Eq.~\eqref{eq:AverageConservatives}), $V_i$ is the effective volume (Eq.~\eqref{eq:EffectiveVolume}), $\overline{\boldsymbol{\mathsf{F}}}_{ij}$ is a numerical flux function, $\overline{\mathbf{Q}}_{ij}$ and $\overline{\dot{\mathbf{r}}}_{ij}$ are an average of the conservative densities and material velocity, respectively, $\mathbf{A}_{ij}$ is the effective interface area (Eq.~\eqref{eq:EffectiveInterfaceArea}) and $\mathbf{S}_{i}$ is a source vector.
The present methodology is not limited to weakly compressible hydrodynamics, but is applicable to any scalar or vectorial balance law that can be cast into Eq.~\eqref{eq:DiscretizedBalanceSystem}, e.g., compressible hydrodynamics~\cite{hopkins2015new, gaburov2011astrophysical}, aeroacoustics~\cite{ramirez2018very}, shallow water equations~\cite{rossi2017well}, and fluctuating hydrodynamics~\cite{serrano2000thermodynamically}.
\par
The discretization of the balance equation (Eq.~\eqref{eq:DiscretizedBalanceSystem}) implies discrete conservation for $\mathbf{S}(\mathbf{Q}) = \mathbf{0}$ as
\begin{equation}
    \tfrac{\mathrm{d}}{\mathrm{d}t}\Big(\sum_{i} \overline{\mathbf{Q}}_{i} V_{i}\Big)
    =
    0,
\end{equation}
see, e.g., Vila~\cite{vila1999particle}, as the effective interface area (Eq.~\eqref{eq:EffectiveInterfaceArea}) is antisymmetric, while the numerical flux function is symmetric by using an appropriate Riemann solver.
\par
A discrete formulation should also conserve angular momentum because this improves accuracy and can stabilize the long-term evolution of highly vortical flows (e.g., Hu and Adams~\cite{hu2006angular}).
However, certain conditions must be satisfied to conserve angular momentum at the discrete level as
\begin{equation}
    \tfrac{\mathrm{d}}{\mathrm{d}t}\Big(\sum_{i}\mathbf{r}_{i}\times m_{i}\overline{\mathbf{v}}_{i}\Big)
    =
    \sum_{i} \underbrace{\Big(\dot{\mathbf{r}}_{i} \times m_{i}\overline{\mathbf{v}}_{i}\Big)}_{\equiv(\dagger)}
    -\frac{1}{2}\sum_{i}\sum_{j}\underbrace{(\mathbf{r}_{i} - \mathbf{r}_{j})\times\Big(\overline{\big[\rho\mathbf{v}\otimes(\mathbf{v} - \dot{\mathbf{r}})+p\mathbf{I}\big]}_{ij}\bullet \mathbf{A}_{ij}\Big)}_{\equiv(\ddagger)}
    =
    0,
\end{equation}
see, e.g., Ivanova et al.~\cite{ivanova2013common}.
The first term $(\dagger)$ of the RHS only vanishes, if the material velocity is equal to the fluid velocity (or zero).
The second term $(\ddagger)$ of the RHS only vanishes if the projection of the comoving numerical flux function onto the normal vector is aligned with the vector connecting the respective material coordinates.
This is the case for, e.g., the standard SPH discretization of the pressure term of Hu and Adams~\cite{hu2006multi, hu2006angular}, but not for a GMFH scheme that utilizes a Riemann solver.
Thus, conservation of angular momentum is not guaranteed for the presented class of GMFH schemes.
\par
Rossi et al.~\cite{rossi2017well} introduced a path-conservative GMFH scheme for hyperbolic systems with non-conservative products.
Their scheme uses a flux-difference formulation that guarantees zeroth-order consistency, but, by design, sacrifices exact conservation.
This exposes a key trade-off in mesh-free discretizations:
A flux-divergence formulation (as used in this work) is exactly conservative but lacks zeroth-order consistency, while a flux-difference formulation achieves consistency at the expense of conservation.
Thus, designing a mesh-free scheme requires choosing between prioritizing conservation and consistency.
Rossi et al.~\cite{rossi2017well} chose the flux-difference form for non-conservative systems such as the shallow water equations and the Baer–Nunziato model for compressible multiphase flows, which are inherently non-conservative.
Recently, interest has grown in applying the flux-difference approach to conservative systems, especially with high-order gradient estimators, e.g.,~\cite{avesani2014new, eiris2023mls, nogueira2016high, ramirez2022arbitrary, ramirez2018very}.
This trend indicates that zeroth-order consistency may, at times, be more important than strict conservation, even when conservation laws are present.
This work prioritizes an exact conservative formulation, as multiphase flows are especially prone to spurious mass and momentum errors that accumulate at material interfaces and can lead to detrimental stability issues.
\par
The definition of the effective volume (Eq.~\eqref{eq:EffectiveVolume}) limits the GMFH scheme to scenarios where the computational volumes are fully supported by neighboring volumes.
Thus, the GMFH scheme cannot handle problems with free surfaces, where the support is truncated by a material interface.
Vila~\cite{vila1999particle} proposed using a transport equation for the effective volume to overcome this constraint, which has become the primary way in SPH to treat free surfaces, see, e.g.,~\cite{sun2017delta,ferrari2009free,michel2023energy,oger2016sph}.
However, the partition of unity property (Eq.~\eqref{eq:PartitionOfUnity}) is not satisfied in that case, and the accumulation of volume errors remains unresolved.
The present methodology is therefore limited to problems where the domain can be fully partitioned into material points without free surfaces; within this class, however, the GMFH scheme handles a wide variety of problems accurately.
\par
The discretization of the balance equations resembles that of a Godunov-type finite volume (FV) scheme.
Eq.~\eqref{eq:DiscretizedBalanceSystem} describes changes in the time rates of average conservative variables, such as, e.g., mass, momentum, or energy, because of the exchange of comoving fluxes through the effective surface of the computational volumes.
The moving-mesh FV scheme of Springel~\cite{springel2010galilean} uses a Voronoi tessellation to compute the true geometric volume and interface area between computational volumes.
The presented GMFH scheme approximates geometric quantities as spatial integrals as no mesh information is available.
Thus, it is at most second-order accurate due to the utilization of a one-point quadrature to approximate the integrals.
\par
The presented GMFH scheme (Eq.~\eqref{eq:DiscretizedBalanceSystem}) is incomplete without the definition of a weighting function and numerical approximations for the effective volume (Eq.~\eqref{eq:EffectiveVolume}) and effective interface area (Eq.~\eqref{eq:EffectiveInterfaceArea}).
In the following, we show that the unified GMFH scheme generalizes a subset of existing methods, including the SPH formulation of Hu and Adams~\cite{hu2006multi} and the WPH formulation of Gaburov and Nitadori~\cite{gaburov2011astrophysical}.
\subsubsection*{Smoothed particle hydrodynamics formulation of Hu and Adams~\cite{hu2006multi}}
The weighting function according to Hu and Adams~\cite{hu2006multi} (or~\cite{hopkins2015new, ivanova2013common, gaburov2011astrophysical}) is
\begin{equation}\label{eq:WeightingFunction}
    \chi_{i}(\mathbf{r})
    \equiv
    \frac{W_{i}(\mathbf{r})}{\sum_{j}W_{j}(\mathbf{r})}
    \equiv
    \frac{W_{i}(\mathbf{r})}{\sigma(\mathbf{r})},
\end{equation}
where $\sigma(\mathbf{r}) = \sum_{j} W_{j}(\mathbf{r})$ is the number density function and
\begin{equation}\label{eq:SmoothingFunction}
    W_{i}(\mathbf{r})
    =
    H^{-d} w\big(\lVert\mathbf{r} - \mathbf{r}_{i}\rVert / H\big)
\end{equation}
is a radial smoothing function in $d$ dimensions.
The smoothing function is centered at $\mathbf{r}_{i}$ with compact support radius $H$, such that $W_{i}(\mathbf{r}) = 0$ for $H \leq \lVert\mathbf{r} - \mathbf{r}_{i}\rVert$  with the property $\int W_{i}(\mathbf{r})\,\mathrm{d}\mathbf{r} = 1$.
The shape of the smoothing function is determined by the dimensionless function $w(r)$, $r \geq 0$, where commonly used smoothing functions are Schoenberg splines~\cite{monaghan1985refined} and Wendland functions~\cite{wendland1995piecewise, wendland2005approximate}, see, e.g., Dehnen et al.~\cite{dehnen2012improving}.
The weighting function~\eqref{eq:WeightingFunction} partitions the $d$-dimensional space into a "smooth" Voronoi tessellation, see, e.g.,~\cite{flekkoy2000foundations, flekkoy1999molecular, serrano2000thermodynamically, hopkins2015new}, where, in contrast to a geometric Voronoi tessellation~\cite{springel2010galilean, gaburro2020high}, the face between adjacent volumes transitions smoothly.
The spatial derivative of the weighting function (Eq.~\eqref{eq:WeightingFunction}) is
\begin{equation}\label{eq:DerivativeOfWeightingFunction}
        \nabla\chi_{i}(\mathbf{r})
        =
        \frac{1}{\sigma(\mathbf{r})^{2}}\sum_{j}\big(W_{j}(\mathbf{r})\nabla W_{i}(\mathbf{r}) - W_{i}(\mathbf{r})\nabla W_{j}(\mathbf{r})\big),
\end{equation}
see, e.g.,~\cite{hu2006multi, hietel2000finite}, where the spatial derivative of the smoothing function is
\begin{equation}
    \nabla W_{i}(\mathbf{r}) = \frac{1}{H^{d + 1}}\frac{\mathbf{r} - \mathbf{r}_{i}}{\lVert \mathbf{r} - \mathbf{r}_{i} \rVert}w'\left(\lVert\mathbf{r} - \mathbf{r}_{i}\rVert/H\right)
\end{equation}
and $w'(r) = \tfrac{\mathrm{d}}{\mathrm{d}r}(w)(r)$ denotes the derivative of the dimensionless function $w(r)$ in Eq.~\eqref{eq:SmoothingFunction}.
The effective volume (Eq.~\eqref{eq:EffectiveVolume}) is approximated using 
\begin{equation}\label{eq:EffectiveVolumeApprox}
    V_{i}
    =
    \int \chi_{i}(\mathbf{r})\,\mathrm{d}\mathbf{r}
    \approx
    \frac{1}{\sigma(\mathbf{r}_{i})}\int W_{i}(\mathbf{r})\,\mathrm{d}\mathbf{r}
    =
    \frac{1}{\sigma(\mathbf{r}_{i})},
\end{equation}
which shows that the effective volume is approximately the inverse number density function.
The effective interface area (Eq.~\eqref{eq:EffectiveInterfaceArea}) using the spatial derivative of the weighting function (Eq.~\eqref{eq:DerivativeOfWeightingFunction}) is
\begin{equation}\label{eq:EffectiveInterParticleSurface}
    \mathbf{A}_{ij}
    \approx
    \left(\frac{1}{\sigma(\mathbf{r}_{i})^{2}}\nabla W_{j}(\mathbf{r}_{i})-\frac{1}{\sigma(\mathbf{r}_{j})^{2}}\nabla W_{i}(\mathbf{r}_{j})\right).
\end{equation}
The effective interface area (Eq.~\eqref{eq:EffectiveInterfaceArea}) recovers the original formulation of Hu and Adams~\cite{hu2006multi} by introducing a uniform support radius as $\nabla W_{i}(\mathbf{r}_{j}) = -\nabla W_{j}(\mathbf{r}_{i})$ for $H(\mathbf{r}_{i}) = H(\mathbf{r}_{j}) = H$.
Thus,
\begin{equation}
    \mathbf{A}_{ij}
    \approx
    -\left(\frac{1}{\sigma(\mathbf{r}_{i})^{2}}+\frac{1}{\sigma(\mathbf{r}_{j})^{2}}\right)\nabla W_{i}(\mathbf{r}_{j})
    \equiv
    \mathbf{N}_{ij}A_{ij},
\end{equation}
where $\mathbf{N}_{ij}W'_{ij} \equiv \nabla W_{i}(\mathbf{r}_{j})$, and the normalized vector $\mathbf{N}_{ij} \equiv \frac{\mathbf{r}_{j} - \mathbf{r}_{i}}{\lVert\mathbf{r}_{j} - \mathbf{r}_{i}\rVert}$ points from the material coordinate $\mathbf{r}_{i}$ to $\mathbf{r}_{j}$.
\par
The presented formalism generalizes the SPH formulation of Hu and Adams~\cite{hu2006multi} by the consistent treatment of an arbitrary material velocity, the utilization of a Riemann solver through the introduction of an appropriate numerical flux function, and the (optional) choice of using variable support radii.
\subsubsection*{Weighted particle hydrodynamics formulation of Gaburov and Nitadori~\cite{gaburov2011astrophysical}}
The weighting function in the formulation of Gaburov and Nitadori~\cite{gaburov2011astrophysical} is equivalent to that of Hu and Adams~\cite{hu2006multi} in Eq.~\eqref{eq:WeightingFunction}.
Consequently, the volume is equivalent to Eq.~\eqref{eq:EffectiveVolumeApprox}.
Instead of using an SPH formulation for the gradient operator, Gaburov and Nitadori~\cite{gaburov2011astrophysical} proposed to use the renormalized gradient formulation of Lanson and Vila~\cite{lanson2008renormalized, lanson2008renormalized2}
\begin{equation}\label{eq:RenormalizedGradientFormulation}
    \langle \nabla\otimes\psi \rangle (\mathbf{r})
    =
    \sum_{j} \big(\psi(\mathbf{r}_{j}) - \psi(\mathbf{r})\big) \otimes \widetilde{\nabla}\chi_{j}(\mathbf{r}),
\end{equation}
where the renormalized gradient of the weighting function is
\begin{equation}\label{eq:GradWeightingFunction}
    \widetilde{\nabla}\chi_{i}(\mathbf{r})
    \equiv
   \boldsymbol{\mathsf{B}}(\mathbf{r})(\mathbf{r}_{i} - \mathbf{r}) \chi_{i}(\mathbf{r}).
\end{equation}
The renormalization matrix $\boldsymbol{\mathsf{B}}\in\mathbf{R}^{d \times d}$ is
\begin{equation}\label{eq:RenormalizationMatrix}
    \boldsymbol{\mathsf{B}}(\mathbf{r}) \equiv (\boldsymbol{\mathsf{E}})^{-1}(\mathbf{r}) \equiv \left(\sum_{j}(\mathbf{r}_{j} - \mathbf{r})\otimes(\mathbf{r}_{j} - \mathbf{r}) \chi_{j}(\mathbf{r})\right)^{-1}.
\end{equation}
The gradient formulation (Eq.~\eqref{eq:RenormalizedGradientFormulation}) is only well-defined as long as the matrix $\boldsymbol{\mathsf{E}}$ is invertible, i.e., not singular, which is, e.g., not the case for an isolated particle without neighboring particles or particles which lie on a perfect line.
We follow Hopkins~\cite{hopkins2015new} and compute the condition number
\begin{equation}\label{eq:ConditionNumber}
    \varkappa_{i} \equiv \frac{1}{d}\sqrt{\big\lVert\boldsymbol{\mathsf{E}}_{i}^{-1}\big\rVert\big\lVert\boldsymbol{\mathsf{E}}_{i}\big\rVert},
\end{equation}
where the respective matrix norm is
\begin{equation}
    \big\lVert\boldsymbol{\mathsf{E}}_{i}\big\rVert
    \equiv
    \sum_{\alpha=1}^{d}\sum_{\beta=1}^{d}\lvert E^{\alpha\beta}_{i}\rvert^{2}.
\end{equation}
When the condition number exceeds a critical condition number $\varkappa_{\mathrm{max}} \equiv 1\cdot10^{2}$, the gradient formulation (Eq.~\eqref{eq:RenormalizedGradientFormulation}) is replaced by the standard SPH gradient of Hu and Adams~\cite{hu2006multi}.
\par
The effective interface area (Eq.~\eqref{eq:EffectiveInterfaceArea}) using the gradient of the weighting function (Eq.~\eqref{eq:GradWeightingFunction}) is
\begin{equation}
    \mathbf{A}_{ij}
    \approx
    \boldsymbol{\mathsf{B}}(\mathbf{r}_{i})(\mathbf{r}_{j} - \mathbf{r}_{i})\frac{W_{j}(\mathbf{r}_{i})}{\sigma(\mathbf{r}_{i})^{2}}
    -
    \boldsymbol{\mathsf{B}}(\mathbf{r}_{j})(\mathbf{r}_{i} - \mathbf{r}_{j})\frac{W_{i}(\mathbf{r}_{j})}{\sigma(\mathbf{r}_{j})^{2}}.
\end{equation}
For uniform support radii, where $W_{i}(\mathbf{r}_{j}) = W_{j}(\mathbf{r}_{i})$ for $H(\mathbf{r}_{i}) = H(\mathbf{r}_{j}) \equiv H$, the effective interface area is
\begin{equation}
    \mathbf{A}_{ij}
    =
    \left(
        \frac{
            \boldsymbol{\mathsf{B}}(\mathbf{r}_{i})
        }{
            \sigma(\mathbf{r}_{i})^{2}
        }
        +
        \frac{
            \boldsymbol{\mathsf{B}}(\mathbf{r}_{j})
        }{
            \sigma(\mathbf{r}_{j})^{2}
        }
    \right)(\mathbf{r}_{j} - \mathbf{r}_{i})W_{i}(\mathbf{r}_{j})
    \equiv
    \mathbf{N}_{ij}A_{ij},
\end{equation}
where $\mathbf{N}_{ij} \equiv \mathbf{A}_{ij} / \lVert\mathbf{A}_{ij}\rVert$ and $A_{ij} \equiv \lVert\mathbf{A}_{ij}\rVert$.
\par
The formulation of Gaburov and Nitadori~\cite{gaburov2011astrophysical} extends the SPH formulation of Hu and Adams~\cite{hu2006multi} by a second-order gradient operator.
This operator accurately reproduces polynomial functions up to the specified order, regardless of particle distribution. However, this improvement requires inverting a small $d \times d$  renormalization matrix for each particle.
Additionally, discrete angular momentum conservation is no longer guaranteed, since the Riemann problem is solved in a direction that does not necessarily align with the line connecting the material coordinates of neighboring volumes.
\subsection{Discretization of the convective flux}\label{sec:ConvectiveFlux}
The computation of the numerical flux function $\overline{\boldsymbol{\mathsf{F}}}_{ij}\equiv\overline{\boldsymbol{\mathsf{F}}}(\mathbf{Q}_{i}, \mathbf{Q}_{j})$ along the interface normal direction $\mathbf{N}_{ij}$ at the interface position 
\begin{equation}
    \mathbf{r}_{ij} \equiv \tfrac{1}{2}(\mathbf{r}_{i} + \mathbf{r}_{j})
\end{equation}
requires the solution of a multidimensional Riemann problem, centered at a frame moving with the interface velocity $\overline{\dot{\mathbf{r}}}_{ij}$.
The numerical flux function solves an exact or approximate Riemann problem and intrinsically includes the necessary numerical dissipation.
In this work we use the HLLC approximate Riemann solver~\cite{toro1994restoration} for its robustness, computational efficiency, and simplicity, see, e.g.,~\cite{toro2009riemann, gaburov2011astrophysical, hu2009hllc,miyoshi2005multi}.
\par
We project the numerical flux function onto the normal plane of the particle interface, such that
\begin{equation}\label{eq:ProjectedNumericalFlux}
    (\overline{\boldsymbol{\mathsf{F}}}_{ij} - \overline{\mathbf{Q}}_{ij}\otimes\overline{\dot{\mathbf{r}}}_{ij})\bullet\mathbf{N}_{ij}
    =
    (\overline{\mathbf{F}}_{ij} - \overline{u}_{ij}\overline{\mathbf{Q}}_{ij})
    \equiv
    \overline{\mathbf{H}}_{ij}.
\end{equation}
Here, $\overline{\mathbf{F}}_{ij}\equiv\overline{\mathbf{F}}(\mathbf{Q}_{i}, \mathbf{Q}_{j})$ and $\overline{\mathbf{H}}_{ij}\equiv\overline{\mathbf{H}}(\mathbf{Q}_{i}, \mathbf{Q}_{j}, \overline{u}_{ij})$ denotes a numerical flux function projected onto the direction of the interface normal direction $\mathbf{N}_{ij}$ moving with the (normal) interface velocity $\overline{u}_{ij} \equiv \overline{\dot{\mathbf{r}}}_{ij}\bullet\mathbf{N}_{ij}$.
\par
We generalize and use $\overline{\mathbf{F}}_{ij}\equiv\overline{\mathbf{F}}(\mathbf{Q}_{ij}^{L}, \mathbf{Q}_{ij}^{R})$ and $\overline{\mathbf{H}}_{ij}\equiv\overline{\mathbf{H}}(\mathbf{Q}^{L}_{ij}, \mathbf{Q}^{R}_{ij}, \overline{u}_{ij})$, i.e., we replace the piecewise constant discontinuous states of the conservative densities by a (yet to be determined) high-order reconstruction of the conservative densities onto the left ($L$) and right ($R$) side of the interface $\mathbf{r}_{ij}$.
The reconstruction of the conservative densities leads to spurious oscillations in the presence of sharp gradients and discontinuities.
In order to avoid oscillatory reconstruction in the vicinity of the material interface, we reconstruct the primitive variables instead.
Using the mapping $\mathcal{Q}:\mathbf{W}\mapsto\mathbf{Q}$ from the primitives to the conservative densities, it follows that $\overline{\mathbf{H}}_{ij}=\overline{\mathbf{H}}(\mathcal{Q}(\mathbf{W}^{L}_{ij}), \mathcal{Q}(\mathbf{W}^{R}_{ij}), \overline{u}_{ij})$.
\subsubsection*{HLLC Riemann solver}
A subsonic (two wave) HLLC Riemann solver moving with the (normal) interface velocity $u_{n}$ computes the numerical flux function for the weakly compressible Euler equations~\eqref{eq:ConservationLaw} from the discontinuous states $\mathbf{Q}_{K}$, $K$ being either the left ($L$) or right ($R$) discontinuous state, such that
\begin{equation}\label{eq:HLLC}
    \overline{\mathbf{H}}^{\mathrm{HLLC}}(\mathbf{Q}_{L},\mathbf{Q}_{R},u_{n})
    \equiv\begin{cases}
    \mathbf{F}(\mathbf{Q}_{L}) + S_{L}(\mathbf{Q}^{\ast}_{L} - \mathbf{Q}_{L}) - u_{n}\mathbf{Q}_{L}^{\ast}
    &
    \mathrm{if}\,u_{n} < S_{M},
    \\
    \mathbf{F}(\mathbf{Q}_{R}) + S_{R}(\mathbf{Q}^{\ast}_{R} - \mathbf{Q}_{R}) - u_{n}\mathbf{Q}_{R}^{\ast}
    &
    \mathrm{if}\,u_{n} \geq S_{M}.
    \end{cases}
\end{equation}
Furthermore, the fluid velocity rotated into the direction of the unit vector $\mathbf{n}$ is $\mathbf{v}\equiv(v_{n}, v_{\tau})$ in two-dimensional space and $\mathbf{v}\equiv(v_{n}, v_{\tau}, v_{\eta})$ in three-dimensional space, where $v_{n}\equiv \mathbf{v} \bullet \mathbf{n}$ denotes the normal component and $v_{\tau}$ and $v_{\eta}$ the components perpendicular to the direction.
The left and right states separated by the contact discontinuity are
\begin{equation}
    \mathbf{Q}_{K}^{\ast}=\rho_{K}\left(\frac{S_{K} - v_{n,K}}{S_{K}-S_{M}}\right)\begin{pmatrix}
    1 \\ S_{M} \\ v_{\tau,K} \\ v_{\eta, K}
    \end{pmatrix}.
\end{equation}
The speed of the contact discontinuity is
\begin{equation}\label{eq:ContactSpeed}
    S_{M} = \frac{p_{R} - p_{L} + \rho_{L} v_{n, L}(S_{L} - v_{n, L}) - \rho_{R} v_{n, R}(S_{R} - v_{n, R})}{\rho_{L}(S_{L} - v_{n, L}) - \rho_{R}(S_{R} - v_{n, R})}.
\end{equation}
The wave speed estimates are in accordance with Davis~\cite{davis1988simplified}
\begin{subequations}
\begin{align}
    S_{L} &= \mathrm{min}(v_{n,L} - c_{0, L}, v_{n, R} - c_{0, R}),
    \\
    S_{R} &= \mathrm{max}(v_{n,L} + c_{0, L}, v_{n, R} + c_{0, R}),
\end{align}
\end{subequations}
\subsubsection*{High-order spatial reconstruction}
A first-order reconstruction of the primitive variables onto the interface $\mathbf{r}_{ij}$ reads
\begin{equation}~\label{eq:FirstOrderReconstruction}
    \mathbf{W}_{ij}^{K}
    =
    \begin{cases}
         \mathbf{W}_{i},
         &
         \mathrm{if}\quad K = \mathrm{L},
         \\
         \mathbf{W}_{j},
         &
         \mathrm{if}\quad K = \mathrm{R}.
    \end{cases}
\end{equation}
The reconstruction of the left and right states using Eq.~\eqref{eq:FirstOrderReconstruction} is straightforward, but introduces an excessive amount of numerical dissipation in smooth flow regions.
Furthermore, it limits the formal order of convergence to first order.
\par
A second-order reconstruction of the primitive variables following Vila~\cite{vila1999particle}, inspired by the MUSCL reconstruction of Van Leer~\cite{vanleer1979towards}, reads
\begin{equation}\label{eq:SecondOrderReconstruction}
    \mathbf{W}_{ij}^{K}
    =
    \begin{cases}
            \mathbf{W}_{i} + \boldsymbol{\alpha}_{i}\langle\nabla\otimes\mathbf{W}\rangle_{i}\bullet(\mathbf{r}_{ij}-\mathbf{r}_{i}),
             &
             \mathrm{if}\quad K = \mathrm{L},
             \\
             \mathbf{W}_{j} + \boldsymbol{\alpha}_{j}\langle\nabla\otimes\mathbf{W}\rangle_{j}\bullet(\mathbf{r}_{ij}-\mathbf{r}_{j}),
             &
             \mathrm{if}\quad K = \mathrm{R},
        \end{cases}
\end{equation}
where we use the renormalized gradient formulation (Eq.~\eqref{eq:RenormalizedGradientFormulation}) to estimate the gradient of the primitive variables.
The reconstruction can be extended beyond second order by employing high-order reconstruction and/or multidimensional WENO reconstruction, see, e.g.,~\cite{nogueira2016high, eiris2023mls, ramirez2022arbitrary, vergnaud2023investigations, avesani2014new}.
Since the flux integral at the interface is approximated using a one-point quadrature (Eq.~\eqref{eq:OnePointQuadrature}), the overall scheme remains second-order accurate irrespective of the reconstruction order.
Nonetheless, these refinements significantly mitigate numerical dissipation.
\par
In regions with steep gradients or poorly distributed particles the gradient approximation (Eq.~\eqref{eq:RenormalizedGradientFormulation}) becomes unreliable.
Thus, slope limiting must be applied to the spatial reconstruction (Eq.~\eqref{eq:SecondOrderReconstruction}) to ensure numerical stability.
A common approach is to limit the gradient (or, equivalently, the slope), such that the reconstructed value is monotonic, i.e., the reconstruction onto the interface does not create new local maxima or minima among the neighboring values.
The multidimensional slope limiter of Balsara~\cite{balsara2004second} as given by Gaburov and Nitadori~\cite{gaburov2011astrophysical} is
\begin{equation}\label{eq:SlopeLimiter}
    \boldsymbol{\alpha}_{i}
    \equiv
    \mathrm{min}\left(1, \mathrm{min}\left(\frac{\mathbf{W}_{i,\mathrm{nbg}}^{\mathrm{max}}-\mathbf{W}_{i}}{\mathbf{W}_{i,\mathrm{mid}}^{\mathrm{max}}-\mathbf{W}_{i}},\frac{\mathbf{W}_{i}-\mathbf{W}_{i,\mathrm{nbg}}^{\mathrm{min}}}{\mathbf{W}_{i}-\mathbf{W}_{i,\mathrm{mid}}^{\mathrm{min}}}\right)\right)\in[0,1],
\end{equation}
where $\mathbf{W}_{i,\mathrm{nbg}}^{\mathrm{min}}$ and $\mathbf{W}_{i,\mathrm{nbg}}^{\mathrm{max}}$ denote the maximum and minimum value of $\mathbf{W}$ within its neighboring particles,
\begin{equation}
    \mathbf{W}_{i,\mathrm{nbg}}^{\mathrm{min}/\mathrm{max}}
    \equiv
    \underset{j}{\mathrm{min}/\mathrm{max}}(\mathbf{W}_{j}),
\end{equation}
$\mathbf{W}_{i,\mathrm{mid}}^{\mathrm{min}}$ and $\mathbf{W}_{i,\mathrm{mid}}^{\mathrm{max}}$ denote the maximum and minimum value of the unlimited reconstruction,
\begin{align}
    \mathbf{W}_{i,\mathrm{mid}}^{\mathrm{min}/\mathrm{max}}
    \equiv
    \mathbf{W}_{i} + \underset{j}{\mathrm{min}/\mathrm{max}}\left(
        \langle\nabla\otimes\mathbf{W}\rangle_{i}\bullet(\mathbf{r}_{ij}-\mathbf{r}_{i})
    \right).
\end{align}
Typically, a pairwise limiting strategy is adopted, see, e.g.,~\cite{michel2023energy,oger2016sph}.
Hopkins~\cite{hopkins2015new} utilized a multidimensional slope limiter and a pairwise slope limiter.
The pairwise limiter does not require an additional particle interaction, but computes a limited slope from the interacting particle pair simultaneously to the spatial reconstruction (Eq.~\eqref{eq:SecondOrderReconstruction}).
This makes the pairwise limiter strategy computationally more efficient, but does not guarantee that the reconstructed value is monotonic.
We did not compare the efficacy of a pairwise limiter to the multidimensional limiter (Eq.~\eqref{eq:SlopeLimiter}), but use the proposed methodology due to its widespread use in unstructured FV schemes~\cite{balsara2004second}, and, while some limiters work better than others, we find that the employed multidimensional limiter (Eq.~\eqref{eq:SlopeLimiter}) works sufficiently well for the problems studied hereinafter.
\subsubsection*{Rotation of the Riemann problem}
The HLLC Riemann solver resolves the contact wave and shear wave in the transverse direction.
Thus, we need to rotate the fluid velocity and material velocity into the normal plane of the Riemann problem to find the correct tangential velocity components.
Euler equations are rotational invariant, see, e.g.,~\cite{ren2003robust, toro2009riemann}.
Thus, the flux function satisfies
\begin{equation}\label{eq:RotationalInvariance}
    \mathbf{F}(\mathbf{Q})
    \equiv
    \boldsymbol{\mathsf{F}}(\mathbf{Q})\bullet\mathbf{N}
    =
    \mathbf{T}^{-1}\mathbf{f}_{1}(\mathbf{T}\mathbf{Q})
    \equiv
    \mathbf{T}^{-1}\mathbf{f}_{1}(\mathbf{Q}')
\end{equation}
as
\begin{subequations}
\begin{align}
    \mathbf{N}
    =
    \sum_{i = 1}^{d} n_{i}\hat{\mathbf{e}}_{i}
    &\overset{(d=2)}{=}
    n_{1}\hat{\mathbf{e}}_{1} + n_{2}\hat{\mathbf{e}}_{2},
    \\
    \boldsymbol{\mathsf{F}}
    =
    \sum_{i = 1}^{d} \mathbf{f}_{i}\otimes\hat{\mathbf{e}}_{i}
    &\overset{(d=2)}{=}
    \mathbf{f}_{1}\otimes\hat{\mathbf{e}}_{1} + \mathbf{f}_{2}\otimes\hat{\mathbf{e}}_{2},
\end{align}
\end{subequations}
where $\hat{\mathbf{e}}_{i}$, $i = 1,\hdots,d$, is a basis vector, $\mathbf{T}$ is a rotation matrix, $\mathbf{T}^{-1}$ is the inverse rotation matrix, and $\mathbf{Q}' \equiv \mathbf{T}\mathbf{Q}$ is the rotated vector of conserved densities.
In $(d=2)$-dimensional space, the rotation matrix and inverse rotation matrix are
\begin{equation}\label{eq:RotationMatrix}
    \mathbf{T}(\mathbf{N}) \equiv
    \begin{pmatrix}
    1 & 0 & 0 \\
    0 & n_{1} & n_{2} \\
    0 & -n_{2} & n_{1} \\
    \end{pmatrix},
    \quad
    \mathbf{T}^{-1}(\mathbf{N}) \equiv
    \begin{pmatrix}
    1 & 0 & 0 \\
    0 & n_{1} & -n_{2} \\
    0 & n_{2} & n_{1} \\
    \end{pmatrix}.
\end{equation}
The property given by Eq.~\eqref{eq:RotationalInvariance} allows us to use a rotated one-dimensional Riemann solver for genuinely multidimensional problems.
Since $\rho$ is invariant under rotation, the rotation matrix $\mathbf{T}$ (Eq.~\eqref{eq:RotationMatrix}) acts identically on the vector of primitive densities, i.e., $\mathbf{W}' \equiv \mathbf{T}\mathbf{W}$ rotates the fluid velocity $\mathbf{v}$ the same way $\mathbf{T}$ rotates the momentum density $\rho\mathbf{v}$ in $\mathbf{Q}' \equiv \mathbf{T}\mathbf{Q}$.
\subsubsection*{Particle interface motion}
Hopkins~\cite{hopkins2015new} defines the material velocity of the interface, such that for the meshless finite-volume (MFV) scheme
\begin{equation}\label{eq:Mfv}
    \overline{u}_{ij}
    \equiv
    \tfrac{1}{2}(\dot{\mathbf{r}}_{i} + \dot{\mathbf{r}}_{j})\bullet\mathbf{N}_{ij},
\end{equation}
and, similar to the approach of Inutsuka~\cite{inutsuka2002reformulation}, for the meshless finite-mass (MFM) scheme
\begin{equation}\label{eq:Mfm}
    \overline{u}_{ij}
    \equiv
    S_M.
\end{equation}
The MFM scheme conserves the mass of individual particles as Eq.~\eqref{eq:Mfm} eliminates mass fluxes between particles.
However, the MFM scheme is only valid, if the material velocity is equal to the fluid velocity, i.e., the motion of the material coordinates is Lagrangian.
The MFV scheme, on the other hand, can be used for an arbitrary material velocity, but mass fluxes between the particles lead to changes in the mass of individual particles even when the material motion is Lagrangian.
\subsubsection*{Algorithm}
We summarize the discretization of the convective flux through the following algorithmic steps for solving the moving Riemann problem at the interface $\mathbf{r}_{ij}$:
\begin{enumerate}
    \item %
        Reconstruct the primitive variables using the second-order reconstruction (Eq.~\eqref{eq:SecondOrderReconstruction}) to obtain the discontinuous primitive states $\mathbf{W}_{ij}^{\mathrm{L}}$ and $\mathbf{W}_{ij}^{\mathrm{R}}$ at the inter-particle interface position.
    \item %
        Transform the discontinuous primitive states $\mathbf{W}_{ij}^{\mathrm{L}}$ and $\mathbf{W}_{ij}^{\mathrm{R}}$ into the rotated Cartesian frame using the rotation matrix $\mathbf{T}_{ij}\equiv\mathbf{T}(\mathbf{N}_{ij})$ (Eq.~\eqref{eq:RotationMatrix}), such that $\mathbf{W}_{ij}^{\mathrm{L}\prime} = \mathbf{T}_{ij}\mathbf{W}_{ij}^{\mathrm{L}}$ and $\mathbf{W}_{ij}^{\mathrm{R}\prime} = \mathbf{T}_{ij}\mathbf{W}_{ij}^{\mathrm{R}}$.
    \item %
        Compute the normal interface velocity $\overline{u}_{ij}$ according to the material velocity formulation:
        \begin{itemize}
            \item %
                For an arbitrary Lagrangian-Eulerian material velocity (or, equivalently, for the MFV scheme), set $\overline{u}_{ij} \equiv \overline{\dot{\mathbf{r}}}_{ij}\bullet\mathbf{N}_{ij}$, where $\overline{\dot{\mathbf{r}}}_{ij}\equiv\tfrac{1}{2}(\dot{\mathbf{r}}_{i}+\dot{\mathbf{r}}_{j})$.
            \item %
                For a genuinely Lagrangian material velocity (or, equivalently, for the MFM scheme), set $\overline{u}_{ij} \equiv S_M$, where $S_M$ is the speed of the contact discontinuity, Eq.~\eqref{eq:ContactSpeed}.
        \end{itemize}
    \item %
        Solve the one-dimensional Riemann problem in the rotated frame using the moving HLLC Riemann solver (Eq.~\eqref{eq:HLLC}) (or any other exact or approximate Riemann solver) to obtain the rotated numerical flux $\overline{\mathbf{H}}_{ij}'=\mathbf{H}(\mathcal{Q}(\mathbf{W}_{ij}^{L\prime}), \mathcal{Q}(\mathbf{W}_{ij}^{R\prime}), \overline{u}_{ij})$.
    \item %
        Transform the rotated numerical flux back onto the original Cartesian frame using the inverse rotation matrix (Eq.~\eqref{eq:RotationMatrix}), such that $\overline{\mathbf{H}}_{ij} = \mathbf{T}^{-1}_{ij}\overline{\mathbf{H}}_{ij}^{\prime}$.
\end{enumerate}
\subsection{Discretization of the arbitrary material velocity}
The GMFH scheme (Eq.~\eqref{eq:DiscretizedBalanceSystem}) belongs to the class of arbitrary Lagrangian-Eulerian (ALE) schemes, where the material velocity $\dot{\mathbf{r}}_{i}$ can be defined arbitrarily. 
An Eulerian scheme is obtained using $\dot{\mathbf{r}}_{i} = \mathbf{0}$, such that
\begin{equation}
    \tfrac{\partial}{\partial t}(\overline{\mathbf{Q}})_{i}
    +
    \frac{1}{V_{i}}\sum_{j} \overline{\boldsymbol{\mathsf{F}}}_{ij}\bullet\mathbf{N}_{ij}A_{ij}
    =
    \mathbf{S}_{i},
\end{equation}
where the material points remain motionless.
A Lagrangian scheme is obtained using $\dot{\mathbf{r}}_{i} = \mathbf{v}(\mathbf{r}_{i})$, such that
\begin{equation}
    \tfrac{\mathrm{d}}{\mathrm{d}t}(\overline{\mathbf{Q}}_{i} V_{i})
    +
    \sum_{j} (\overline{\boldsymbol{\mathsf{F}}}_{ij} - \overline{\mathbf{Q}}_{ij}\otimes\overline{\mathbf{v}}_{ij})\bullet\mathbf{N}_{ij}A_{ij}
    =
    \mathbf{S}_{i} V_{i},
\end{equation}
where the material points move with the fluid velocity.
\par
An Eulerian scheme inherits many limitations of grid-based schemes\footnote{Several of these limitations are not exclusive to Eulerian schemes and depend on the specific discretization.}, such as the challenging representation of complex geometries with moving boundaries, discretization errors associated with the convective term, lack of Galilean invariance in the discrete equations, and potential artifacts due to numerical dissipation at material interfaces unless specific interface-capturing techniques are applied, such as the level-set method~\cite{gibou2018review} or diffuse-interface methods~\cite{johnsen2006implementation,coralic2014finite,paula2023robust}.
On the other hand, maintaining a purely Lagrangian formulation often leads to highly anisotropic particle distributions, as highlighted by several authors~\cite{oger2016sph, adami2013transport, antuono2021delta}.
This increases numerical errors~\cite{litvinov2015towards} and compromises the method's robustness~\cite{adami2013transport}.
\par
A quasi-Lagrangian material velocity is defined such that
\begin{equation}\label{eq:QuasiLagrangianMaterialVelocity}
    \dot{\mathbf{r}}_{i} \equiv \mathbf{v}_{i} + \delta\mathbf{v}_{i},
\end{equation}
where the particle-slip (or particle-perturbation) velocity $\delta\mathbf{v}_{i} = \dot{\mathbf{r}}_{i} - \mathbf{v}_{i}$ is small, i.e., $\lVert\delta\mathbf{v}_{i}\rVert \ll \lVert\mathbf{v}_{i}\rVert$, and the Lagrangian nature of the scheme is preserved as much as possible.
The particle-slip velocity should include a relaxation mechanism that improves particle distribution and, for consistency, converges to the fluid velocity as spatial resolution increases.
\par
Several different formulations exist for defining an appropriate material velocity which produces a quasi-Lagrangian scheme, see, e.g.,~\cite{zhang2017generalized, oger2016sph, adami2013transport, antuono2021delta, michel2022particle, lind2012incompressible}.
At its core, the particle-slip velocity utilizes
\begin{itemize}[label={--}]
    \item %
        an inconsistent gradient approximation of a constant field, e.g., by a constant background pressure in Adami et al.~\cite{adami2013transport} or by particle concentration in Lind et al.~\cite{lind2012incompressible}, as a relaxation mechanism to perturb the particle trajectories into the direction of insufficient support,
    \item %
        a scaling coefficient which decreases the influence of the particle-slip velocity with increasing spatial resolution, e.g., the smoothing length in Oger et al.~\cite{oger2016sph} or timestep size in Adami et al.~\cite{adami2013transport}, and
    \item %
        a characteristic velocity that scales the particle-slip velocity appropriately.
\end{itemize}
We refer to the work of Michel et al.~\cite{michel2022particle} for a comprehensive comparison of different particle-slip velocities.
They showed that, while different formulations exist, the particle-slip velocity can be cast into the following generic formulation, where
\begin{equation}\label{eq:PerturbationVelocity}
    \widetilde{\delta\mathbf{v}_{i}}
    \equiv
    -U^{\mathrm{char}}d^{\mathrm{char}}\langle\nabla\otimes 1 \rangle^{\mathrm{SPH}}(\mathbf{r}_{i}),
\end{equation}
and
\begin{equation}\label{eq:LimitDisplacementVelocity}
    \delta\mathbf{v}_{i}
    \equiv
    \alpha
    \cdot
    \begin{cases}
        \displaystyle\widetilde{\delta\mathbf{v}_{i}}
        &
        \mathrm{if}
        \quad
        \lVert \widetilde{\delta\mathbf{v}_{i}} \rVert < U^{\mathrm{lim}},
        \\
        \displaystyle U^{\mathrm{lim}}\widetilde{\delta\mathbf{v}_{i}}\big/\lVert\widetilde{\delta\mathbf{v}_{i}}\rVert
        &
        \mathrm{else}.
    \end{cases}
\end{equation}
The definition of the characteristic velocity $U^\mathrm{char}$, characteristic length $d^\mathrm{char}$, and maximum particle-slip magnitude $U^{\mathrm{lim}}$ determines how aggressively and accurately the particle-slip velocity is able to improve the particle distribution.
The limiting of the particle-slip velocity (Eq.~\eqref{eq:LimitDisplacementVelocity}), as well as introducing the constant $\alpha$ to adjust the magnitude of the non-Lagrangian particle displacement, has been introduced to prevent excessive particle movement that can have a detrimental effect on stability.
We use $\alpha = 0.5$ in this work.
The gradient of unity is used to find a vector field, that points from regions of low particle concentration to high particle concentration, taking advantage of the inconsistent standard SPH gradient discretization.
\par
In this work, we use a particle-perturbation velocity formulation that is similar to that of Michel et al.~\cite{michel2022particle}, which, in comparison to other formulations, is Galilean invariant, rotationally invariant, and consistent.
Thus,
\begin{equation}
    d^\mathrm{char} \equiv H,
\end{equation}
\begin{equation}
    U^\mathrm{char}
    \equiv
    \max_{j}\left\lvert(\mathbf{v}_{j} - \mathbf{v}_{i})\bullet\frac{\mathbf{r}_{j} - \mathbf{r}_{i}}{\lVert\mathbf{r}_{j} - \mathbf{r}_{i}\rVert}\right\rvert,
\end{equation}
and
\begin{equation}
    U^{\mathrm{lim}}\equiv U^\mathrm{char}.
\end{equation}
We use the SPH gradient formulation of Hu and Adams~\cite{hu2006multi}
\begin{equation}\label{eq:StandardSPHGradient}
    \langle\nabla\otimes 1 \rangle^{\mathrm{SPH}}(\mathbf{r}_{i})
    \equiv
    \sigma(\mathbf{r}_{i}) \sum_{j}
    \left(
        \frac{1}{\sigma^{2}(\mathbf{r}_{i})}
        +
        \frac{1}{\sigma^{2}(\mathbf{r}_{j})}
    \right)\nabla W_{j}(\mathbf{r}_{i})\big|_{H = 2 \Delta x_{0}},
\end{equation}
where the support radius of the standard kernel gradient is modulated to mitigate the detrimental effect of pairing instabilities using the characteristic inter-particle distance $\Delta x_{0}$.
\subsection{Material interface treatment}\label{sec:MaterialInterfaceTreatment}
We consider $N_{\mathrm{C}}$ immiscible fluid phases, which are labeled $\alpha = 1, \hdots, N_{\mathrm{C}}$.
The color (index) function of Hu and Adams~\cite{hu2006multi}
\begin{equation}\label{eq:ColorFunction}
    C^{\alpha}(\mathbf{r})
    \equiv
    \begin{cases}
        1 & \mathrm{if}\quad \mathbf{r}\text{ is in phase }\alpha, \\
        0 & \mathrm{if}\quad \mathbf{r}\text{ is not in phase }\alpha,
    \end{cases}
\end{equation}
is used to distinguish material coordinates of different fluid phases.
Every material coordinate belongs to exactly one phase, so $\sum_{\alpha} C^{\alpha}_{i} = 1$, and since phases are carried by Lagrangian material coordinates, $\dot{C}^{\alpha}_{i} = 0$.
The color gradient between the phases $\alpha$ and $\beta$ is approximated using the SPH gradient formulation of Hu and Adams~\cite{hu2006multi}, such that for $\alpha\neq\beta$
\begin{equation}\label{eq:ColorGradient}
    \langle\nabla\otimes C\rangle^{\alpha\beta}_{i}
    =
    \sigma_{i} \sum_{j}
    \left(
        \frac{C^{\beta}_{i}}{\sigma^{2}_{i}}
        +
        \frac{C^{\beta}_{j}}{\sigma^{2}_{j}}
    \right)\nabla W_{j}(\mathbf{r}_{i}),
\end{equation}
where $C^{\beta}_{i}=0$.
The geometric normal vector at the material coordinate $\mathbf{r}_{i}$ from phase $\alpha$ to every other phase is $\widehat{\mathbf{N}}^{\alpha}_{i} \equiv \mathbf{N}^{\alpha}_{i} /\lVert\mathbf{N}^{\alpha}_{i}\rVert$, where
\begin{equation}\label{eq:NormalVector}
    \mathbf{N}^{\alpha}_{i} 
    \equiv
    \sum_{\substack{\beta=1 \\ \alpha\neq\beta}}^{N_{\mathrm{C}}}\langle\nabla\otimes C\rangle^{\alpha\beta}_{i}.
\end{equation}
The color function (Eq.~\eqref{eq:ColorFunction}), the color gradient (Eq.~\eqref{eq:ColorGradient}), and the normal vector (Eq.~\eqref{eq:NormalVector}) hold for an arbitrary number of fluid phases, which is a distinct advantage of the formalism outlined in this work.
\par
For two-phase problems ($N_{\mathrm{C}}=2$), we characterize the density contrast between the phases by the density ratio
\begin{equation}\label{eq:DensityRatio}
    \Theta \equiv \frac{\rho_{0,\ell}}{\rho_{0,h}},
\end{equation}
where $\rho_{0,\ell}$ and $\rho_{0,h}$ are the reference densities (Eq.~\eqref{eq:EquationOfState}) of the light ($\ell$) and heavy ($h$) phase, respectively.
A small $\Theta$ corresponds to a large density contrast between the phases, whereas $\Theta \to 1$ recovers phases of equal density.
\subsubsection*{One-sided renormalized gradient formulation}
The reconstruction of primitive variables onto the inter-particle interface position requires approximating their gradients, see Eq.~\eqref{eq:SecondOrderReconstruction}.
In multiphase problems, the density field has a discontinuity at the material interface.
As a result, a standard gradient approximation yields erroneous results near these regions.
We propose to
\begin{itemize}[label={--}]
    \item %
        compute the gradient of velocity symmetrically as described by the standard renormalized gradient formulation of Lanson and Vila~\cite{lanson2008renormalized, lanson2008renormalized2} in Eq.~\eqref{eq:RenormalizedGradientFormulation}, and
    \item %
        compute the gradient of density in a one-sided manner, i.e., biased towards the respective phase of the material coordinate.
\end{itemize}
The one-sided renormalized gradient formulation for an arbitrary property is
\begin{equation}\label{eq:OneSidedGradientRenormalization}
    \langle \nabla\otimes\psi \rangle^{\alpha}_{i}
    =
     \boldsymbol{\mathsf{B}}^{\alpha}(\mathbf{r}_{i})
     \sum_{\substack{j \\ C_{i}^{\alpha} = C_{j}^{\alpha} = 1}} \big(\psi(\mathbf{r}_{j}) - \psi(\mathbf{r}_{i})\big) \otimes (\mathbf{r}_{j} - \mathbf{r}_{i})\chi_{j}(\mathbf{r}_{i}),
\end{equation}
where
\begin{equation}\label{eq:OneSidedRenormalizationMatrix}
    \boldsymbol{\mathsf{B}}^{\alpha}(\mathbf{r}_{i})
    \equiv
    \left(\sum_{\substack{j \\ C_{i}^{\alpha} = C_{j}^{\alpha} = 1}}(\mathbf{r}_{j} - \mathbf{r}_{i})\otimes(\mathbf{r}_{j} - \mathbf{r}_{i}) \chi_{j}(\mathbf{r}_{i})\right)^{-1}.
\end{equation}
The one-sided renormalization matrix (Eq.~\eqref{eq:OneSidedRenormalizationMatrix}) becomes singular when a particle is completely surrounded by a different phase, with no neighboring particles of the same phase.
To address this, we compute the condition number (Eq.~\eqref{eq:ConditionNumber}) of the one-sided renormalization matrix (Eq.~\eqref{eq:OneSidedRenormalizationMatrix}) and revert to a one-sided SPH gradient formulation when it exceeds the critical threshold.
For an isolated particle in a different phase, the one-sided SPH gradient yields zero, resulting in first-order reconstruction of the respective primitive variable.
\subsubsection*{Material interface motion}
The quasi-Lagrangian material velocity formulation (Eq.~\eqref{eq:QuasiLagrangianMaterialVelocity}) effectively mitigates the formation of numerical voids and fragmentation due to tensile instability and pairing instability.
However, it does not respect the natural Lagrangian motion of the material interface, because the interface kinematics is determined only by the normal fluid velocity at the interface~\cite{tryggvason2011direct}.
The kinematic motion of the material interface is naturally respected by using 
\begin{equation}\label{eq:MaterialInterfaceMotion}
    \dot{\mathbf{r}}_{i} \equiv \mathbf{v}_{i} +
    \begin{cases}
        (\mathbf{I} - \widehat{\mathbf{N}}^{\alpha}_{i}\otimes\widehat{\mathbf{N}}^{\alpha}_{i})\bullet\delta\mathbf{v}_{i}
        &
        \mathrm{if}\quad \mathbf{N}_{i}^{\alpha} \neq \mathbf{0},
        \\
        \delta\mathbf{v}_{i}
        &
        \mathrm{else},
    \end{cases}
\end{equation}
see, e.g.,~\cite{lind2012incompressible,sun2017delta,mokos2016multiphase}, where $\mathbf{N}_{i}^{\alpha} \neq \mathbf{0}$ (Eq.~\eqref{eq:NormalVector}) indicates that particle $i$ has at least one neighbor in a different phase, i.e., $i$ lies at the material interface, and the quasi-Lagrangian material velocity (Eq.~\eqref{eq:QuasiLagrangianMaterialVelocity}) is used everywhere apart from this region.
Near the material interface, only the tangential particle-slip velocity is considered such that the normal material velocity is not altered, preserving the natural motion of the interface.
\par
We have experimented with
    \begin{equation}
        \dot{\mathbf{r}}_{i} \equiv \mathbf{v}_{i} +
        \begin{cases}
            \mathbf{0}
            &
            \mathrm{if}\quad \mathbf{N}_{i}^{\alpha} \neq \mathbf{0},
            \\
            \delta\mathbf{v}_{i}
            &
            \mathrm{else}.
        \end{cases}
    \end{equation}
    We find that for problems where the tangential velocity is large compared to the normal velocity at the material interface, such as the Kelvin-Helmholtz and Rayleigh-Taylor instabilities, the particle distribution near the interface is poor.
    The tangential regularization of Eq.~\eqref{eq:MaterialInterfaceMotion} significantly improves the isotropy of the particle distribution in these cases.
\par
Rezavand et al.~\cite{rezavand2020weakly} and Mokos et al.~\cite{mokos2016multiphase} use a different strategy to maintain the Lagrangian kinematics of the material interface:
    The material velocity in the light phase is a quasi-Lagrangian formalism similar to Eq.~\eqref{eq:QuasiLagrangianMaterialVelocity}, while in the heavy phase, a Lagrangian material velocity is utilized.
    This strategy respects the natural motion of the interface as long as the density ratio $\Theta$ is sufficiently small.
    This ensures that the underlying assumption, that the material interface is dominated by the Lagrangian motion of the heavy phase, holds.
    The drawback of this strategy is that it breaks down for moderate values of $\Theta$.
    It also remains unclear how small $\Theta$ must be for this approach to remain valid.
\subsubsection*{Particle interface motion}
The individual mass of particles varies as a result of mass fluxes between neighboring particles.
This can introduce numerical instabilities when $\Theta$ is small due to mass fluxes from the heavy to the light phase.
We propose to use the following interface velocity of the moving Riemann problem, where
\begin{equation}\label{eq:ParticleInterfaceVelocity}
    \overline{u}_{ij}
    \equiv
    \begin{cases}
        S_{M}
        &
        \mathrm{if}\quad C_{i}^{\alpha} \neq C_{j}^{\alpha},
        \\
        \tfrac{1}{2}(\dot{\mathbf{r}}_{i} + \dot{\mathbf{r}}_{j})\bullet\mathbf{N}_{ij}
        &
        \mathrm{else}.
    \end{cases}
\end{equation}
This strategy enhances numerical stability while preserving mass exchange dynamics for same-phase particle interactions, since setting $\overline{u}_{ij} \equiv S_{M}$ across the material interface eliminates mass flux there entirely.
\subsection{Time integration scheme}
An explicit Drift-Kick-Drift (DKD) integration scheme for the material coordinates and the conservative variables, see, e.g.,~\cite{springel2005cosmological, monaghan2005smoothed}, is given by
\begin{subequations}\label{eq:DKD}
\begin{align}
    \mathbf{r}^{(n+1/2)}_{i}
    &=
    \mathbf{r}^{(n)}_{i} + \tfrac{1}{2} \Delta t^{(n)}\cdot\tfrac{\mathrm{d}}{\mathrm{d}t}(\mathbf{r})^{(n)}_{i},
    \\
    \label{eq:kick}
    (\overline{\mathbf{Q}}V)^{(n+1)}_{i} 
    &=
    (\overline{\mathbf{Q}}V)^{(n)}_{i} + \Delta t^{(n)}\cdot\tfrac{\mathrm{d}}{\mathrm{d}t}(\overline{\mathbf{Q}}V)^{(n)}_{i},
    \\
    \mathbf{r}^{(n+1)}_{i}
    &=
    \mathbf{r}^{(n+1/2)}_{i} + \tfrac{1}{2} \Delta t^{(n)}\cdot\tfrac{\mathrm{d}}{\mathrm{d}t}(\mathbf{r})^{(n+1)}_{i},
\end{align}
\end{subequations}
where $\Delta t^{(n)}$ denotes the variable timestep (given in Eq.~\eqref{eq:CflCriterion}) at the integration step $n$.
\par
Following Gaburov and Nitadori~\cite{gaburov2011astrophysical}, we use a multistage Runge-Kutta (RK) integration scheme in this work to advance the conservatives from the integration step $n$ to $n + 1$.
Thus, we replace the single-stage kick operator (Eq.~\eqref{eq:kick}) with a two-stage second-order accurate RK (RK22) or a three-stage third-order accurate RK (RK33) time integration scheme.
We refer to the combination of DKD and RK time integration as DKDRK22 and DKDRK33.
The DKDRK22 time integration scheme is given by
\begin{subequations}\label{eq:DKDRK22}
\begin{align}
    \mathbf{r}^{(n+1/2)}_{i}
    &=
    \mathbf{r}^{(n)}_{i} + \tfrac{1}{2} \Delta t^{(n)}\cdot\tfrac{\mathrm{d}}{\mathrm{d}t}(\mathbf{r})^{(n)}_{i},
    \\
    (\overline{\mathbf{Q}}V)^{(\ast)}_{i}
    &=
    (\overline{\mathbf{Q}}V)^{(n)}_{i} + \Delta t^{(n)} \cdot \tfrac{\mathrm{d}}{\mathrm{d}t}(\overline{\mathbf{Q}}V)^{(n)}_{i},
    \\
    (\overline{\mathbf{Q}}V)^{(n + 1)}_{i}
    &=
    \tfrac{1}{2}(\overline{\mathbf{Q}}V)^{(n)}_{i} + \tfrac{1}{2}\left((\overline{\mathbf{Q}}V)^{(\ast)}_{i} + \Delta t^{(n)} \cdot \tfrac{\mathrm{d}}{\mathrm{d}t}(\overline{\mathbf{Q}}V)^{(\ast)}_{i}\right),
    \\
    \mathbf{r}^{(n+1)}_{i}
    &=
    \mathbf{r}^{(n+1/2)}_{i} + \tfrac{1}{2} \Delta t^{(n)}\cdot\tfrac{\mathrm{d}}{\mathrm{d}t}(\mathbf{r})^{(n+1)}_{i},
\end{align}
\end{subequations}
and the DKDRK33 time integration scheme is given by
\begin{subequations}\label{eq:DKDRK33}
\begin{align}
    \mathbf{r}^{(n+1/2)}_{i}
    &=
    \mathbf{r}^{(n)}_{i} + \tfrac{1}{2} \Delta t^{(n)}\cdot\tfrac{\mathrm{d}}{\mathrm{d}t}(\mathbf{r})^{(n)}_{i},
    \\
    (\overline{\mathbf{Q}}V)^{(\ast)}_{i}
    &=
    (\overline{\mathbf{Q}}V)^{(n)}_{i} + \Delta t^{(n)} \cdot \tfrac{\mathrm{d}}{\mathrm{d}t}(\overline{\mathbf{Q}}V)^{(n)}_{i},
    \\
    (\overline{\mathbf{Q}}V)^{(\ast\ast)}_{i}
    &=
    \tfrac{3}{4}(\overline{\mathbf{Q}}V)^{(n)}_{i} + \tfrac{1}{4}\left((\overline{\mathbf{Q}}V)^{(\ast)}_{i} + \Delta t^{(n)} \cdot \tfrac{\mathrm{d}}{\mathrm{d}t}(\overline{\mathbf{Q}}V)^{(\ast)}_{i}\right),
    \\
    (\overline{\mathbf{Q}}V)^{(n + 1)}_{i}
    &=
    \tfrac{1}{3}(\overline{\mathbf{Q}}V)^{(n)}_{i} + \tfrac{2}{3}\left((\overline{\mathbf{Q}}V)^{(\ast\ast)}_{i} + \Delta t^{(n)} \cdot \tfrac{\mathrm{d}}{\mathrm{d}t}(\overline{\mathbf{Q}}V)^{(\ast\ast)}_{i}\right),
    \\
    \mathbf{r}^{(n+1)}_{i}
    &=
    \mathbf{r}^{(n+1/2)}_{i} + \tfrac{1}{2} \Delta t^{(n)}\cdot\tfrac{\mathrm{d}}{\mathrm{d}t}(\mathbf{r})^{(n+1)}_{i},
\end{align}
\end{subequations}
where $(\ast)$ and $(\ast\ast)$ denote the intermediate stages.
Geometric quantities such as the effective particle volume, renormalization matrix, and color gradients depend only on the material coordinates and do not need to be recomputed between RK stages.
\par
The maximum timestep is limited by the CFL criterion
\begin{equation}\label{eq:CflCriterion}
    \Delta t = \mathrm{CFL} \cdot \min_{i}\left(\frac{\ell_{i}}{c_{0, i} + \lVert\mathbf{v}-\dot{\mathbf{r}}\rVert_{i}}\right),
\end{equation}
where, for clarity, we drop the superscript denoting the integration step $n$.
For DKDRK22 (Eq.~\eqref{eq:DKDRK22}) and DKDRK33 (Eq.~\eqref{eq:DKDRK33}), $\mathrm{CFL} < 1$ for stability, see, e.g.,~\cite{gottlieb2005high, ketcheson2008highly}.
We use the characteristic particle size $\ell_{i} \equiv (V_{i} / C_{d})^{1/d}$, where $C_d$ is the volume of a unit sphere in $d$-dimensional space, see, e.g.,~\cite{gaburov2011astrophysical,springel2010galilean}.
\par
We note that neither the single-stage kick operator (Eq.~\eqref{eq:DKD}) nor the multi-stage kick operator (Eq.~\eqref{eq:DKDRK22} and Eq.~\eqref{eq:DKDRK33}) is symplectic, in contrast to, e.g., a symplectic and time-reversible leapfrog integrator as given by Monaghan~\cite{monaghan2005smoothed}.
Symplecticity is a structural property of integrators for separable Hamiltonian systems with conservative, position-dependent forces.
This property is exploited by Hamiltonian SPH formulations that integrate their equations of motion with a symplectic leapfrog scheme, see, e.g.,~\cite{grenier2009hamiltonian, monaghan2005smoothed}.
However, in this work, the kick operator advances conservative fluid variables using numerical fluxes obtained from the solution of Riemann problems at the particle interfaces (Eq.~\eqref{eq:ProjectedNumericalFlux}).
Therefore, the underlying semi-discrete system is dissipative rather than Hamiltonian, and no symplectic structure exists at the spatial-discretization level for any single- or multistage kick operator to preserve.
Instead, following Gaburov and Nitadori~\cite{gaburov2011astrophysical}, we choose a RK kick operator for improved accuracy and stability.
\subsection{Boundary conditions}
We use the ghost-particle technique in this work to describe boundary conditions, see, e.g.,~\cite{adami2012generalized, morris1997modeling}.
Ghost particles occupy the boundary regions to ensure that fluid particles in the vicinity of the boundary are fully supported.
In this work, ghost particles remain at fixed material coordinates and do not carry their own conservative or primitive states.
\par
We rewrite the flux exchange between neighboring particles in Eq.~\eqref{eq:DiscretizedBalanceSystem} using Eq.~\eqref{eq:ProjectedNumericalFlux}, such that
\begin{align}
    \sum_{j\in\mathcal{F}\cup\mathcal{G}} \overline{\mathbf{H}}_{ij}A_{ij}
    =
    \sum_{j\in\mathcal{F}}\overline{\mathbf{H}}_{ij}A_{ij}
    +
    \sum_{j\in\mathcal{G}}\overline{\mathbf{H}}(\mathcal{Q}(\mathbf{W}^{L}_{ij}), \mathcal{Q}(\mathbf{W}_{ij}^{R\ast}), \overline{u}_{ij})A_{ij}.
\end{align}
Here, $\mathcal{F}$ denotes the fluid region and $\mathcal{G}$ the ghost region.
When a fluid particle interacts with a ghost particle, we solve a one-sided Riemann problem similar to classical finite-volume schemes, see, e.g.,~\cite{toro2009riemann, leveque1988high}, as described by Zhang et al.~\cite{zhang2017weakly} for mesh-free methods.
The left state corresponds to the reconstruction from the fluid phase, while the right state is an augmented fluid state that satisfies the relevant boundary conditions.
\par
For non-reflecting boundary conditions, such as outflow boundaries, the augmented right state of the Riemann problem is given by
\begin{equation}
    \mathbf{W}_{ij}^{R\ast} = \begin{pmatrix}
        \rho_{ij}^{L}
        \\[0.2em]
        \mathbf{v}_{ij}^{L} 
    \end{pmatrix}.
\end{equation}
This approach is valid only when the fluid velocity is small, as ghost particles remain at fixed material coordinates in this work.
\par
For reflecting boundary conditions, such as wall boundaries, we use
\begin{equation}
    \mathbf{W}_{ij}^{R\ast} = \begin{pmatrix}
        \rho_{ij}^{L}
        \\[0.2em]
        \mathbf{v}_{ij}^{L} - (\mathbf{v}_{ij}^{L}\bullet\widehat{\mathbf{N}}_{i}^{\alpha})\widehat{\mathbf{N}}_{i}^{\alpha}
    \end{pmatrix}.
\end{equation}
The treatment of boundary conditions closely follows the treatment of a material interface, with a few straightforward modifications:
The second-order reconstruction of the primitive fluid states onto the particle interface is carried out using the one-sided renormalized gradient formulation (Eq.~\eqref{eq:OneSidedGradientRenormalization}) for density and velocity.
The material velocity of fluid particles near the boundary and the material velocity of the particle interface follow the same multiphase formulation given in Eq.~\eqref{eq:MaterialInterfaceMotion} and Eq.~\eqref{eq:ParticleInterfaceVelocity}.
\section{Numerical tests}\label{sec:NumericalTests}
If not stated otherwise, we initialize the particles on a Cartesian lattice using the initial lattice spacing $\Delta x_{0}$.
We exclusively use the cubic kernel function as given by Dehnen et al.~\cite{dehnen2012improving} with a relative compact support radius $H = 2.8\Delta x$.
All simulation results were generated using double precision accuracy if not stated otherwise, see Appendix~\ref{sec:FloatingPointAccuracy} for a detailed discussion and comparison between single and double precision computation.
The semi-discrete governing equations are integrated in time using the DKDRK33 (Eq.~\eqref{eq:DKDRK33}) integrator with $\mathrm{CFL} = 0.85$.
If not stated otherwise, we use the same speed of sound for the light fluid and the heavy fluid.
If not stated otherwise, the force density field is zero, i.e., $\mathbf{f} \equiv \mathbf{0}$.
We denote the spatial coordinates as $\mathbf{x} = (x, y)$ and the fluid velocity as $\mathbf{v} = (u, v)$ in two-dimensional space.
We emphasize, that throughout this work we do not utilize any explicit form of physical or artificial dissipation mechanism but rely on the discretization scheme itself for numerical stability.
\subsubsection*{The role of artificial speed of sound}
The speed of sound in weakly compressible hydrodynamics is a problem-dependent parameter that limits density fluctuation to 1\%.
As a general rule of thumb, we use
\begin{equation}
    c_{0} = 10\cdot\mathrm{max}\left(\max_{i}\lVert\mathbf{v}_{i}\rVert,\sqrt{\frac{\Delta p_{i}}{\rho_{0}}}\right),
\end{equation}
see, e.g.,~\cite{sun2017delta, colagrossi2003numerical, grenier2009hamiltonian}.
For scenarios, where the heavy fluid strongly compresses the light fluid, e.g., the Rayleigh-Taylor instability, we increase the speed of sound in the light phase as discussed by Colagrossi and Landrini~\cite{colagrossi2003numerical}.
\subsubsection*{The role of background pressure}
The continuous governing equations (Eq.~\eqref{eq:ConservationLaw}) are invariant to an arbitrary background pressure in the EOS (Eq.~\eqref{eq:EquationOfState}).
However, a conservative GMFH scheme is not zeroth-order consistent, and negative pressures (or tension) can trigger tensile instability.
A suitably chosen background pressure can mitigate this effect.
\par
In this work, we use $p_b = 0$ because the natural relaxation mechanism of the quasi-Lagrangian material velocity mitigates the detrimental effect of tensile instability.
The only exception is the Kelvin-Helmholtz instability.
The tangential projection of the particle-slip velocity is not sufficient to prevent numerical voids near the material interface.
In this case, we found that a small background pressure $p_b = 0.01 \cdot c_{0}^{2} \Theta$ prevents void formation in the interfacial region without deteriorating the solution.
The elimination of the need for a background pressure for the Kelvin-Helmholtz instability will be addressed in future work.
\subsection{Single-phase problems}
\subsubsection{Taylor-Green vortex}
The Taylor-Green vortex (TGV) is a classical hydrodynamical test problem for studying the properties and accuracy of numerical schemes, see, e.g.,~\cite{hu2006multi, hu2006angular, hu2007incompressible, hu2015sph, adami2013transport, zhang2017weakly, zhang2025towards}.
In this work we study the inviscid TGV to investigate the stability and dissipation mechanism of the proposed GMFH scheme in the limit of infinite Reynolds number.
\par
The fluid is initialized in a periodic domain with $0.0 \leq x \leq 1.0$ and $0.0 \leq y \leq 1.0$.
The initial conditions for the primitive variables are
\begin{subequations}
    \begin{align}
        \rho(\mathbf{x}, t = 0)
        &=
        \rho_{0} + \tfrac{1}{4}\rho_{0}\big(\mathrm{cos}(2kx)+\mathrm{cos}(2ky)\big)/c_{0}^{2},
        \\
        u(\mathbf{x}, t = 0)
        &=
        -U\,\mathrm{cos}(kx)\,\mathrm{sin}(ky),
        \\
        v(\mathbf{x}, t = 0)
        &=
        U\,\mathrm{sin}(kx)\,\mathrm{cos}(ky),
    \end{align}
\end{subequations}
where $k = 2\pi$ is the wave number and $U=1.0$ is a characteristic velocity.
The artificial speed of sound is $c_{0} = 10.0$ and the reference density is $\rho_{0} = 1.0$.
\par
The particle distribution and fluid velocity are given in Fig.~\ref{fig:TgvParticleDistribution}.
We compare the Lagrangian MFM and Lagrangian MFV schemes of Hopkins~\cite{hopkins2015new} with an MFV scheme using a quasi-Lagrangian material velocity according to Eq.~\eqref{eq:QuasiLagrangianMaterialVelocity}.
We find that the particle relaxation mechanism significantly improves the particle distribution and, consequently, the accuracy of the GMFH scheme~\cite{litvinov2015towards}.
We did not compare our implementation with the variable smoothing-length approach of Hopkins~\cite{hopkins2015new}.
A variable smoothing-length implementation could further improve accuracy by compensating for locally poor particle distributions in weakly compressible flows.
We proceed to use the quasi-Lagrangian MFV scheme with the naturally incorporated particle relaxation mechanism unless stated otherwise.
\par
\begin{figure}[h]
    \centering
    \includegraphics[page=1]{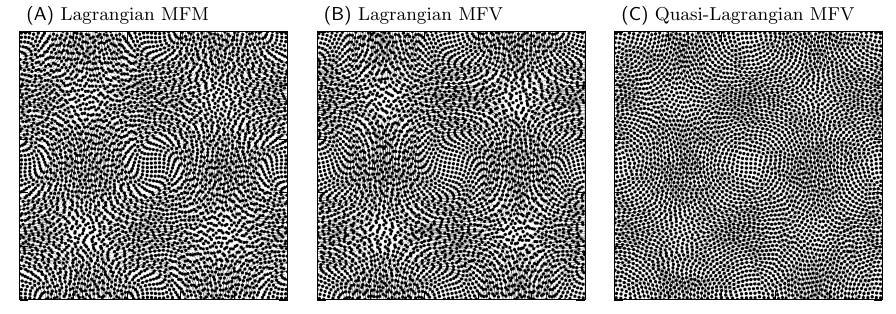}
    \caption{
        The particle distribution and fluid velocity of the TGV using \textsf{(A)} Lagrangian MFM, \textsf{(B)} Lagrangian MFV, and \textsf{(C)} quasi-Lagrangian MFV with particle relaxation at $t=1.0$ and $\Delta x_{0} = 1/64$.
    }
    \label{fig:TgvParticleDistribution}
\end{figure}
The relative mass and relative linear momentum are given in Fig.~\ref{fig:TgvFloatingPointAccuracy}.
We compare single-precision with double-precision accuracy to demonstrate that global mass and global linear momentum are conserved within the implementation's floating-point accuracy.
\par
\begin{figure}[h]
    \centering
    \includegraphics[page=1]{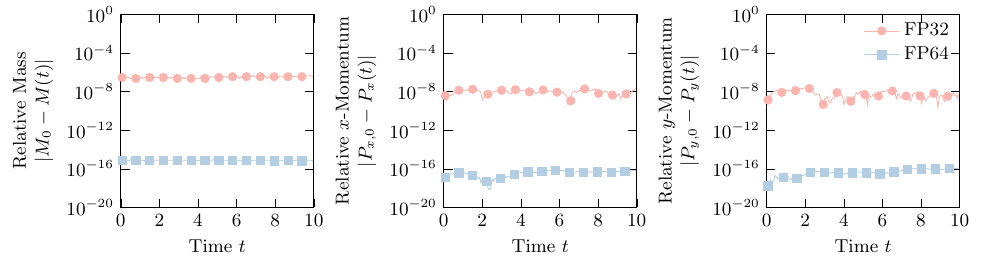}
    \caption{
        The relative mass and relative linear momentum of the TGV using single-precision accuracy (\raisebox{-0.2ex}{\scalebox{1.3}{\textcolor{Pastel1-1}{$\bullet$}}}) and double-precision accuracy (\scalebox{0.8}{\textcolor{Pastel1-2}{$\blacksquare$}}) at $\Delta x_{0} = 1/64$.
    }
    \label{fig:TgvFloatingPointAccuracy}
\end{figure}
The maximum velocity, normalized kinetic energy, angular momentum error, and volume error are given in Fig.~\ref{fig:TgvKineticEnergy}.
We compare the solution for an initially Cartesian and relaxed particle packing at different particle resolutions.
The relaxed particle distribution is obtained from the final distribution of the initially Cartesian particle packing.
The kinetic energy (and entropy) of the incompressible TGV without physical dissipation is a conserved quantity, but this does not translate to the discrete level, see, e.g.,~\cite{honein2004higher, pirozzoli2010generalized, morinishi2010skew}, unless a specific discretization is used.
We find that the dissipation of kinetic energy and maximum velocity decreases with increased resolution.
The initially Cartesian particle packing introduces spurious oscillations in kinetic energy, which do not appear when starting from a relaxed configuration.
We find that angular momentum conservation improves with increasing particle resolution.
The error in total volume remains relatively constant across different particle resolutions.
The error can be further reduced by increasing the support radius of the smoothing function, though this also increases dissipation.
\begin{figure}[h!]
    \centering
    \includegraphics[page=1]{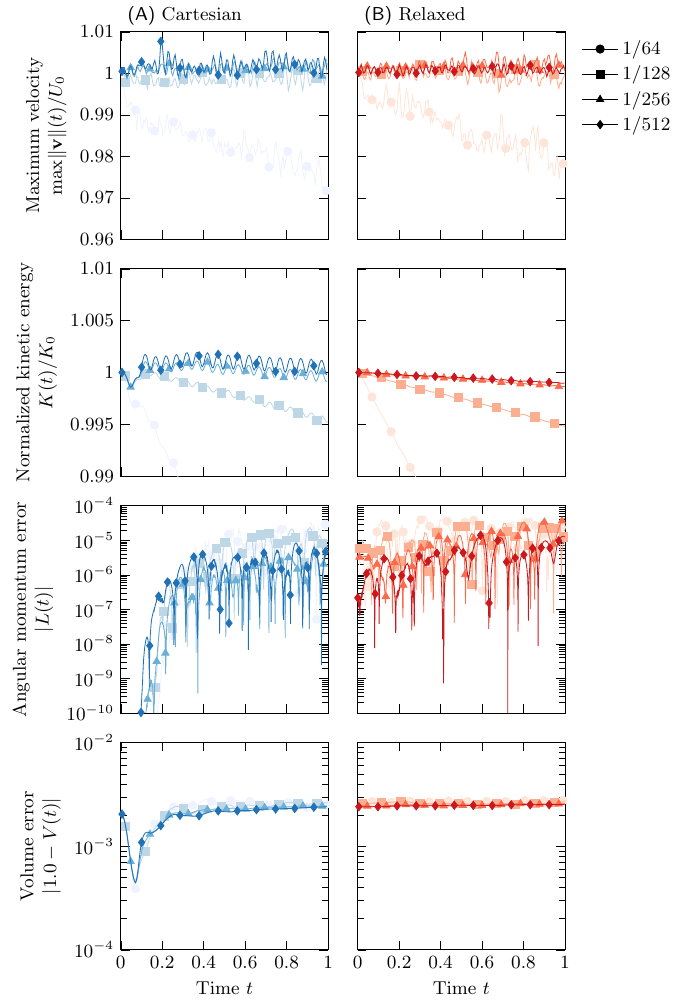}
    \caption{
        The maximum velocity, normalized kinetic energy, angular momentum error, and volume error of the TGV for an initially \textsf{(A)} Cartesian particle packing and \textsf{(B)} relaxed particle packing at $\Delta x_{0} = 1/64$ (\raisebox{-0.2ex}{\scalebox{1.3}{$\bullet$}}), $\Delta x_{0} = 1/128$ (\scalebox{0.8}{$\blacksquare$}), $\Delta x_{0} = 1/256$ ($\blacktriangle$), and $\Delta x_{0} = 1/512$ ($\blacklozenge$).
    }
    \label{fig:TgvKineticEnergy}
\end{figure}
\par
The fluid velocity as vector plot for a long-term simulation is given in Fig.~\ref{fig:TgvLongTerm}.
We observe similar behavior to the two-dimensional decay of turbulence, where the array of vortices merges into a vortex pair with opposite strength, see, e.g., Hu and Adams~\cite{hu2015sph}, due to the presence of a small (but non-vanishing) numerical dissipation.
When increasing the particle resolution, the initial vortex configuration is maintained for a longer time period due to the diminishing effect of numerical dissipation.
\begin{figure}[h]
    \centering
    \includegraphics[page=1]{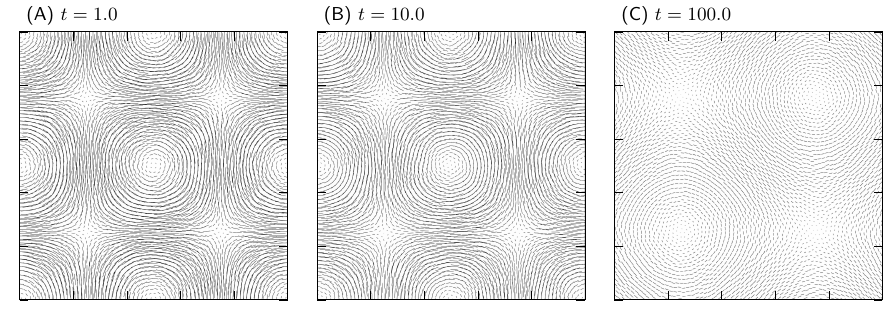}
    \caption{
        The fluid velocity as vector plot of the TGV at \textsf{(A)} $t = 1.0$, \textsf{(B)} $t = 10.0$, and \textsf{(C)} $t=100.0$ and $\Delta x_{0} = 1/64$.
    }
    \label{fig:TgvLongTerm}
\end{figure}
\subsection{Multiphase problems}
\subsubsection{Material interface advection}
The material interface advection (MIA) describes the advection of an initially square fluid patch in a uniform velocity field.
Researchers have used this problem in various forms, see, e.g.,~\cite{hopkins2015new, hess2010particle, cha2010kelvin, saitoh2013density, hopkins2013general}, to show that numerical dissipation does not excessively smear the material interface.
In this work, we use this simple but challenging problem to show that the proposed methodology maintains hydrodynamic equilibrium, preserves the Lagrangian nature of the material interface, and remains stable even for low density ratios.
\par
The fluid is initialized in a periodic domain with $0.0 \leq x \leq 1.0$ and $0.0 \leq y \leq 1.0$.
The initial conditions for the primitive variables are
\begin{subequations}
    \begin{align}
        \rho(\mathbf{x}, t = 0)
        &=
        \begin{cases}
            \rho_{0, h} 
            &
            \mathrm{if}\quad 0.3 \leq x \leq 0.7\quad\mathrm{and}\quad 0.3 \leq y \leq 0.7,
            \\
            \rho_{0,\ell}
            &
            \mathrm{else},
        \end{cases}
        \\
        u(\mathbf{x}, t = 0)
        &=
        1.0,
        \\
        v(\mathbf{x}, t = 0)
        &=
        1.0,
    \end{align}
\end{subequations}
The artificial speed of sound is $c_{0,h} = c_{0,\ell} = 14.2$ for both phases.
The reference density is $\rho_{0,\ell} = \Theta$ and $\rho_{0,h} = 1.0$ in the light and in the heavy phase, respectively.
We investigate the density ratio $\Theta = 1:10^{16}$.
\par
The particle distribution is given in Fig.~\ref{fig:MaterialInterfaceAdvection}.
The position and shape of the material interface are captured accurately at all investigated time instances.
The scheme remains stable even at extremely low density ratios around machine precision without numerical instabilities because of the careful design of the multiphase treatment and limiting strategy.
We find that (not shown here) the MIA problem using our multiphase GMFH scheme is very similar to the solution obtained from an MFM scheme.
The problem is trivial for the Lagrangian MFM scheme as it is solved exactly to machine precision, see, e.g., Hopkins~\cite{hopkins2015new}.
However, as shown for the TGV problem, weakly compressible flows with complex features require a particle relaxation scheme to achieve better accuracy and stability.
We stress that this problem requires a Galilean-invariant particle relaxation scheme, as otherwise the relaxation mechanism leads to large errors in the long-time evolution of the interface position and shape.
\begin{figure}[h]
    \centering
    \includegraphics[page=1]{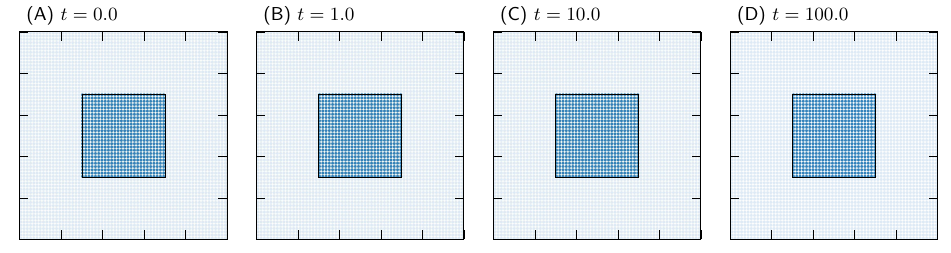}
    \caption{
        The particle distribution of the light phase ($\textcolor{Blues3-1}{\bullet}$) and heavy phase ($\textcolor{Blues3-3}{\bullet}$) for the MIA problem at $\mathsf{(A)}$ $t = 0.0$, $\mathsf{(B)}$ $t = 1.0$, $\mathsf{(C)}$ $t = 10.0$, and $\mathsf{(D)}$ $t = 100.0$ and $\Delta x_{0} = 1 / 64$.
        The black line represents the exact position of the material interface at the given time instances.
    }
    \label{fig:MaterialInterfaceAdvection}
\end{figure}
\subsubsection{Surface gravity waves}
The surface gravity wave (SGW) problem describes the propagation of waves along the material interface between a light and a heavy fluid in a gravitational field.
It is one of the few multiphase flow problems for which linear theory provides an exact and non-trivial analytical solution~\cite{kundu2024fluid,choudhuri1998physics,lamb1924hydrodynamics}, making it an effective benchmark for numerical methods.
It also evaluates a method's ability to capture moving interfaces and identify numerical artifacts like spurious currents and instabilities.
The SGW problem has previously been investigated in the context of SPH methods by Monaghan and Rafiee~\cite{monaghan2013simple}.
\par
The fluid particles are initialized in a domain with $-0.5 \leq x \leq 0.5$ and $-1.0 \leq y \leq 1.0$.
We use periodic boundary conditions between the left and right boundary, and non-reflecting boundary conditions at the top and bottom boundary.
The external force density (or acceleration) field is $\mathbf{f} = (0, -g)^{\mathrm{T}}$, where $g = 1$.
The initial conditions are
\begin{subequations}
    \begin{align}
        \rho(\mathbf{x}, t = 0)
        &=
        \begin{cases}
            \rho_{0,\ell} + (\rho_{0,\ell}  - y \rho_{0,\ell}) / c_{0,\ell}^{2},
            &
            \mathrm{if}\quad y \geq 0
            \\
            \rho_{0,h} + (\rho_{0,\ell} - y \rho_{0,h}) / c_{0, h}^{2},
            &
            \mathrm{else},
        \end{cases}
        \\
        u(\mathbf{x}, t = 0)
        &=
        \begin{cases}
            \displaystyle+0.01\sin(k x) \frac{\cosh(k (y - 1.0))}{\sinh(k)}
            &
            \mathrm{if}\quad y \geq 0
            \\
            \displaystyle-0.01\sin(k x) \frac{\cosh(k (y + 1.0))}{\sinh(k)}
            &
            \mathrm{else},
        \end{cases}
        \\
        v(\mathbf{x}, t = 0)
        &=
        \begin{cases}
            \displaystyle-0.01\cos(k x) \frac{\sinh(k (y - 1.0))}{\sinh(k)}
            &
            \mathrm{if}\quad y \geq 0
            \\
            \displaystyle+0.01\cos(k x) \frac{\sinh(k (y + 1.0))}{\sinh(k)}
            &
            \mathrm{else},
        \end{cases}
    \end{align}
\end{subequations}
where $k = 2\pi$ is the wave number.
We initialize a perturbation of the velocity field instead of initializing the displacement of the material interface, which requires carefully relaxing the initial particle distribution.
The initial density profile is an approximation to hydrostatic equilibrium.
The exact frequency for the SGW problem is, according to, e.g., Choudhuri~\cite{choudhuri1998physics},
\begin{equation}\label{eq:SurfaceGravityWavesFrequency}
    \omega_\mathrm{SGW} = \sqrt{g k \frac{\rho_{0, h} - \rho_{0, \ell}}{\rho_{0, h} + \rho_{0, \ell}}}.
\end{equation}
We normalize time using $\tau\equiv t / T$, where $T\equiv 2\pi/\omega_\mathrm{SGW}$, such that the material interface is in its initial position every full unit in normalized time.
The reference density is $\rho_{0,\ell} = \Theta$ and $\rho_{0,h} = 1.0$ in the light and in the heavy phase, respectively.
The artificial speed of sound is $c_{0,h} = c_{0,\ell} = 10.0$ for all density ratios.
\par
The particle distributions of the light and heavy phases for various density ratios are shown in Fig.~\ref{fig:SurfaceGravityWavesParticles}.
The material interface remains sharp across all density ratios; no artificial mixing or particle penetration is observed.
Moreover, the particle relaxation mechanism does not cause any artificial breakup of the material interface, even for extremely low density ratios.
The absence of a relaxation mechanism normal to the material interface results in a less isotropic particle distribution in the interface region than in the bulk, especially for the heavy phase.
Despite that, we observe no numerical instabilities in the interface region.
The numerical dissipation reduces the maximum interface deflection over time as expected.
\begin{figure}[h!]
    \centering
    \includegraphics[page=1]{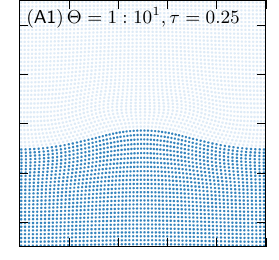}\hspace{-0.35cm}
    \includegraphics[page=2]{figures/surface-gravity-wave_2d_particles.pdf}\hspace{-0.35cm}
    \includegraphics[page=3]{figures/surface-gravity-wave_2d_particles.pdf}
    \includegraphics[page=4]{figures/surface-gravity-wave_2d_particles.pdf}\hspace{-0.35cm}
    \includegraphics[page=5]{figures/surface-gravity-wave_2d_particles.pdf}\hspace{-0.35cm}
    \includegraphics[page=6]{figures/surface-gravity-wave_2d_particles.pdf}
    \includegraphics[page=7]{figures/surface-gravity-wave_2d_particles.pdf}\hspace{-0.35cm}
    \includegraphics[page=8]{figures/surface-gravity-wave_2d_particles.pdf}\hspace{-0.35cm}
    \includegraphics[page=9]{figures/surface-gravity-wave_2d_particles.pdf}
    \includegraphics[page=10]{figures/surface-gravity-wave_2d_particles.pdf}\hspace{-0.35cm}
    \includegraphics[page=11]{figures/surface-gravity-wave_2d_particles.pdf}\hspace{-0.35cm}
    \includegraphics[page=12]{figures/surface-gravity-wave_2d_particles.pdf}
    \caption{
        The particle distribution of the light phase ($\textcolor{Blues3-1}{\bullet}$) and heavy phase ($\textcolor{Blues3-3}{\bullet}$) for the SGW problem for the density ratio (\textsf{A}) $\Theta = 1:10$, (\textsf{B}) $\Theta = 1:10^{2}$, (\textsf{C}) $\Theta = 1:10^{3}$, and (\textsf{D}) $\Theta = 1:10^{6}$ at particle resolution $\Delta x_{0} = 1/64$ and normalized time (\textsf{1}) $\tau = 0.25$, (\textsf{2}) $\tau = 1.25$, and (\textsf{3}) $\tau = 9.25$.
    }
    \label{fig:SurfaceGravityWavesParticles}
\end{figure}
\par
The normalized kinetic energy for various density ratios is given in Fig.~\ref{fig:SurfaceGravityWaves}.
We observe, that the dissipation of kinetic energy due to numerical dissipation decreases with increased particle resolution as expected.
The dissipation of kinetic energy for lower density ratios is less pronounced than for higher density ratios.
\begin{figure}[h!]
    \centering
    \includegraphics{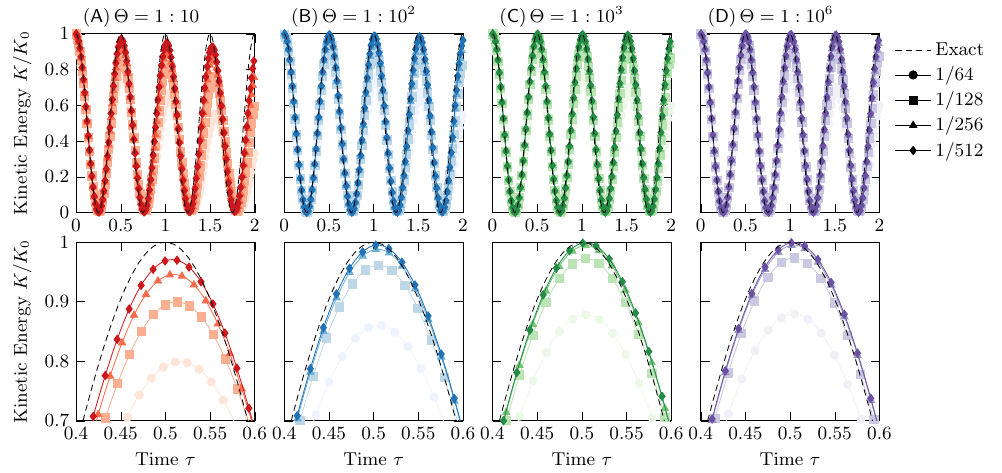}
    \caption{
        The normalized kinetic energy $K / K_{0}$ over normalized time $\tau$ for the SGW problem for the density ratio (\textsf{A}) $\Theta = 1:10$, (\textsf{B}) $\Theta = 1:10^{2}$, (\textsf{C}) $\Theta = 1:10^{3}$, and (\textsf{D}) $\Theta = 1:10^{6}$ at particle resolution $\Delta x_{0} = 1/64$ (\raisebox{-0.2ex}{\scalebox{1.3}{$\bullet$}}), $\Delta x_{0} = 1/128$ (\scalebox{0.8}{$\blacksquare$}), $\Delta x_{0} = 1/256$ ($\blacktriangle$), and $\Delta x_{0} = 1/512$ ($\blacklozenge$).
        The black dashed line represents the exact normalized kinetic energy.
    }
    \label{fig:SurfaceGravityWaves}
\end{figure}
\subsubsection{Kelvin-Helmholtz instability}
The Kelvin-Helmholtz instability (KHI) is a fundamental hydrodynamic instability that occurs when two adjacent fluids move relative to each other.
The velocity shear causes initially small perturbations at the material interface to grow into characteristic vortices.
The instability evolves through two stages: A linear stage, where small perturbations grow exponentially to form coherent vortex structures, and a nonlinear stage, where these vortices interact and merge into larger-scale structures through vortex pairing and breakdown.
Without stabilizing mechanisms such as gravitational stratification, surface tension, or viscosity, the interface is unconditionally unstable because perturbations of any wavelength will grow over time.
The KHI arises in many engineering applications and natural phenomena, including atmospheric flows~\cite{klaassen1985evolution}, oceanic mixing~\cite{smyth2012ocean}, and planetary magnetospheres~\cite{johnson2014kelvin}.
The KHI has been studied extensively within the astrophysics community using SPH~\cite{yue2015numerical, cha2010kelvin, valcke2010kelvin, junk2010modelling}.
However, accurate simulation of KHI with classical SPH remains challenging due to two main limitations:
Density discontinuities at material interfaces generate spurious interfacial pressures that act as artificial surface tension. Excessive numerical dissipation also suppresses the growth of perturbation modes.
We demonstrate that the present methodology successfully captures the physics of this challenging mechanism.
\par
The fluid is initialized in a periodic domain with $0.0 \leq x \leq 1.0$ and $0.0 \leq y \leq 1.0$.
The initial conditions for the primitive variables are
\begin{subequations}
    \begin{align}
        \rho(\mathbf{x}, t = 0)
        &=
        \begin{cases}
            \rho_{0,h}
            &
            \mathrm{if}\quad 0.25 \leq y \leq 0.75,
            \\
            \rho_{0,\ell}
            &
            \mathrm{else},
        \end{cases}
        \\
        u(\mathbf{x}, t = 0)
        &=
        \begin{cases}
            +U
            &
            \mathrm{if}\quad 0.25 \leq y \leq 0.75,
            \\
            -U
            &
            \mathrm{else},
        \end{cases}
        \\
        v(\mathbf{x}, t = 0)
        &=
        0.01\sin(k x),
    \end{align}
\end{subequations}
where $k = 4\pi$ is the perturbation wave number and $U = 1.0$ is a characteristic velocity.
The linear growth rate of the KHI is, see, e.g.,~\cite{choudhuri1998physics, chandrasekhar2013hydrodynamic,wang2010combined},
\begin{equation}
    \omega_\mathrm{KHI}
    =
    2 U \frac{\sqrt{\rho_{0,\ell}\rho_{0, h}}}{\rho_{0,\ell}+\rho_{0,h}}k.
\end{equation}
We normalize time using $T = 2\pi/\omega_{\mathrm{KHI}}$.
The reference density is $\rho_{0,\ell} = \Theta$ and $\rho_{0,h} = 1.0$ in the light and in the heavy phase, respectively.
The artificial speed of sound is $c_{0} = 15.0$ for all density ratios.
\par
The particle distributions of the light and heavy phases for different density ratios are shown in Fig.~\ref{fig:KelvinHelmholtzParticles}.
The particle distribution remains isotropic even in the vicinity of the material interface for all investigated density ratios, indicating stable numerical behavior.
Small wavelength disturbances are visible due to the initial lattice structure of the particles in the early stages of the KHI.
For the equal density ratio, the initial perturbations grow until vortex roll-up is developed, revealing the characteristic pattern of the KHI.
In later stages, the instability develops into larger, more complex vortical structures.
For the moderate density ratio, vortex formation is comparable to that at the equal density ratio at early times.
At later times, the vortex dynamics differ from those at equal density ratios because of the material interface dynamics.
At the lowest density ratio, the initial perturbations grow stronger than at the equal and moderate density ratios.
At later times, the evolution of the vortices is less coherent than at higher density ratios.
As the dynamics become violent due to the momentum difference between the two phases, we observe strong phase mixing and fragmentation of the heavier phase into filaments.
This behavior reflects the absence of any physical stabilizing mechanisms, such as, e.g., stratification, surface tension, or viscosity.
\begin{figure}[h!]
    \centering
    \includegraphics{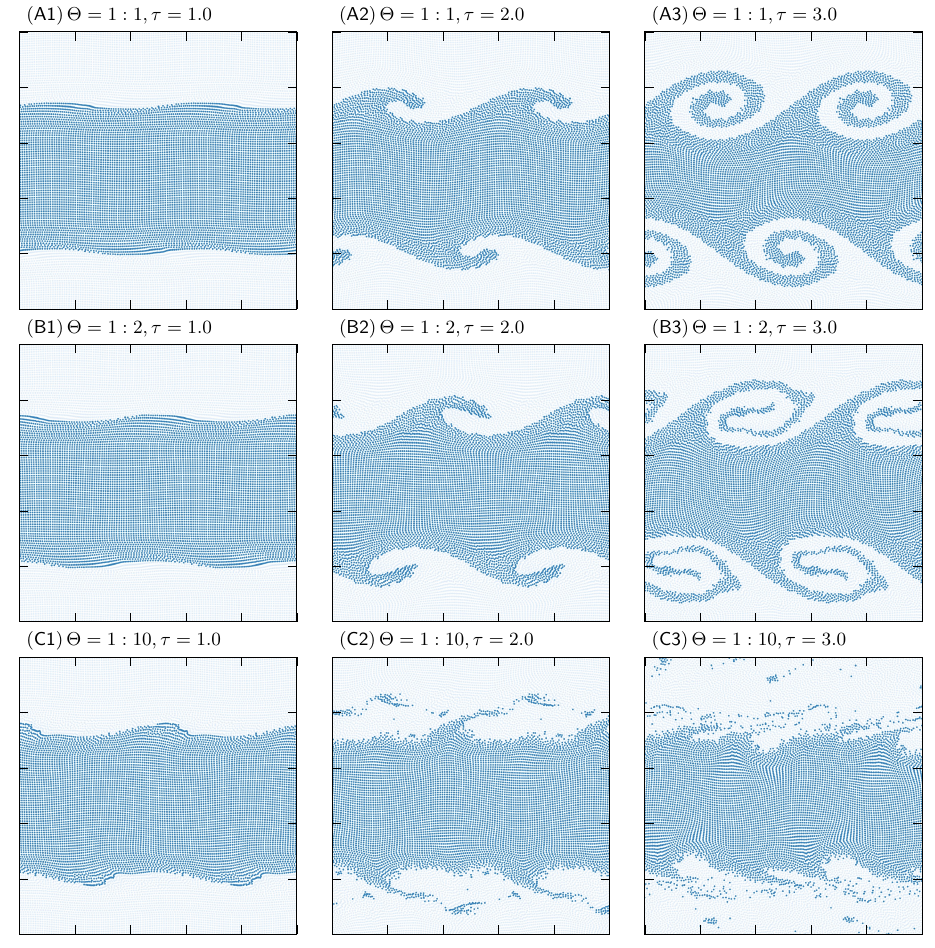}
    \caption{
        The particle distribution of the light phase ($\textcolor{Blues3-1}{\bullet}$) and heavy phase ($\textcolor{Blues3-3}{\bullet}$) for the KHI for the density ratio (\textsf{A}) $\Theta = 1:1$, (\textsf{B}) $\Theta = 1:2$, and (\textsf{C}) $\Theta = 1:10$, at the particle resolution $\Delta x_{0} = 1/128$ and normalized time (\textsf{1}) $\tau = 1.0$, (\textsf{2}) $\tau = 2.0$, and (\textsf{3}) $\tau = 3.0$.
    }
    \label{fig:KelvinHelmholtzParticles}
\end{figure}
\par
The mode amplitude for different density ratios and particle resolutions is shown in Fig.~\ref{fig:KelvinHelmholtzMode}.
We compute the linear-mode amplitude for wave number $k = 4\pi$ using a discrete convolution as proposed in McNally et al.~\cite{mcnally2012wellposed} and compare our results to the exact relative amplitude in the linear growth regime.
We observe that at higher resolution, all density ratios converge to the mode amplitude of the theoretical solution in the linear growth regime.
\begin{figure}[h!]
    \centering
    \includegraphics[page=1]{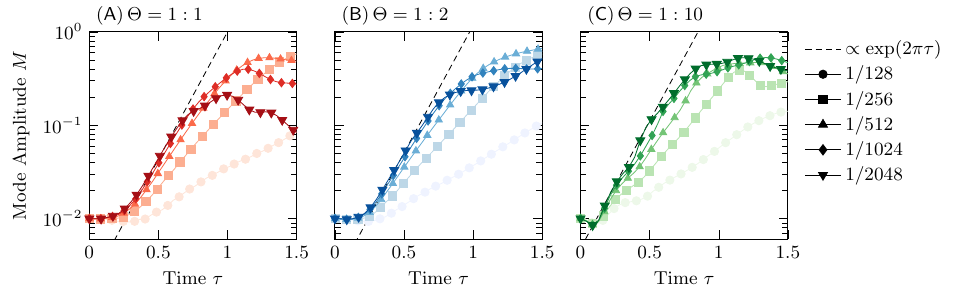}
    \caption{
        The linear mode amplitude $M$ of the $k = 4 \pi$ perturbation mode over non-dimensional time $\tau$ for the KHI for $\mathsf{(A)}$ $\Theta = 1 : 1$, $\mathsf{(B)}$ $\Theta = 1 : 2$ and $\mathsf{(C)}$ $\Theta = 1 : 10$ at particle resolution $\Delta x_{0} = 1/128$ (\raisebox{-0.2ex}{\scalebox{1.3}{$\bullet$}}), $\Delta x_{0} = 1/256$ (\scalebox{0.8}{$\blacksquare$}), $\Delta x_{0} = 1/512$ ($\blacktriangle$), $\Delta x_{0} = 1/1024$ ($\blacklozenge$), and $\Delta x_{0} = 1/2048$ ($\blacktriangledown$).
        The black dashed line (--\,--) represents the exact relative amplitude in the linear growth regime.
    }
    \label{fig:KelvinHelmholtzMode}
\end{figure}
\par
The smoothed color function for different density ratios at the finest particle resolution is shown in Fig.~\ref{fig:KelvinHelmholtzColor}, which highlights the evolution of primary and secondary instabilities in a highly resolved GMFH simulation.
The smoothed color function is obtained by kernel interpolation of the particles' color function onto an even higher-resolved grid.
Compared to the particle distribution in Fig.~\ref{fig:KelvinHelmholtzParticles}, initial disturbances grow faster in a highly resolved simulation because of diminishing numerical viscosity, as indicated by the linear mode amplitude in Fig.~\ref{fig:KelvinHelmholtzMode}.
For an equal density ratio, initial roll-up is followed by vortex pairing and merging that enhance fluid-phase mixing. Decreasing the density ratio shifts the dynamics towards secondary instabilities localized at the material interface and, at the lowest ratio, towards fragmentation of the heavy phase into small-scale filaments.
\begin{figure}[h!]
    \centering
    \includegraphics{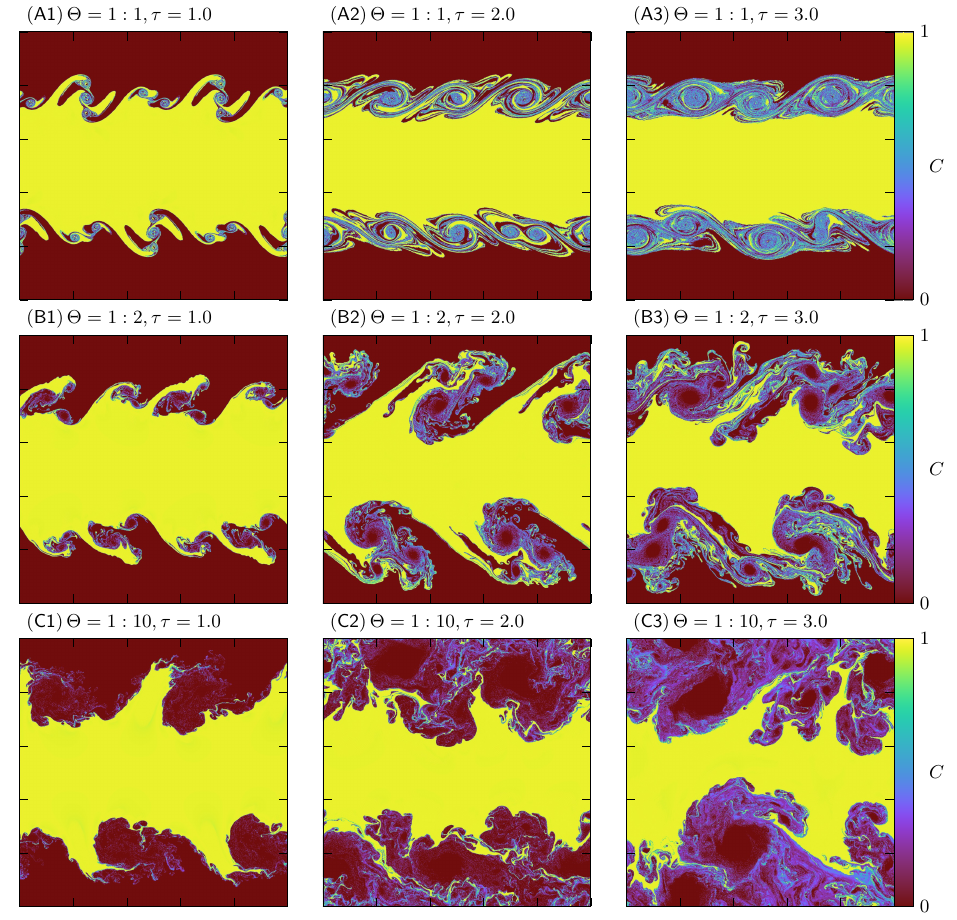}
    \caption{
        The smoothed color function $C$ for the KHI for the density ratios $\mathsf{(A)}$ $\Theta = 1 : 1$, $\mathsf{(B)}$ $\Theta = 1 : 2$ and $\mathsf{(C)}$ $\Theta = 1 : 10$ at non-dimensional time $\mathsf{(A)}$ $\tau = 1.0$, $\mathsf{(B)}$ $\tau = 2.0$, and $\mathsf{(C)}$ $\tau = 3.0$ for particle resolution $\Delta x_{0} = 1/2048$.
    }
    \label{fig:KelvinHelmholtzColor}
\end{figure}
\subsubsection{Rayleigh-Taylor instability}
The Rayleigh-Taylor instability (RTI) is a fundamental hydrodynamic instability that occurs at the material interface when a heavier fluid is accelerated into a lighter fluid by gravity.
The RTI is unconditionally unstable in the absence of surface tension and viscosity as any perturbation of the interface grows exponentially, leading to rising bubbles of light fluid and descending spikes of heavy fluid.
This often leads to secondary instabilities, such as Kelvin-Helmholtz instabilities.
The RTI is a ubiquitous fluid phenomenon observed in nature and engineering applications. It reduces efficiency in inertial confinement fusion (ICF) due to premature mixing of fuel and ablator material~\cite{betti2016inertial, zhou2025instabilities, atzeni2005fluid}, leads to mushroom-shaped structures in the earth's atmosphere and oceans~\cite{zhou2017rayleigh, zhou2017rayleigh2}, and contributes to efficient fuel-air mixing in combustion processes~\cite{zhou2017rayleigh}.
The RTI has been studied using GMFH schemes before, see, e.g.,~\cite{monaghan2013simple, adami2013transport, hopkins2015new}.
In this work, we use the proposed methodology to systematically investigate the RTI across increasingly challenging density ratios at different particle resolutions.
\par
The fluid is initialized in a domain with $-0.5 \leq x \leq 0.5$ and $-1.0 \leq y \leq 1.0$.
We use periodic boundary conditions between the left and right boundary, and non-reflecting boundary conditions at the top and bottom boundary.
The external force density (or acceleration) field is $\mathbf{f} = (0, -g)^{\mathrm{T}}$, where $g = 1$.
The initial conditions for the primitive variables are
\begin{subequations}
    \begin{align}
        \rho(\mathbf{x}, t = 0)
        &=
        \begin{cases}
            \rho_{0,h} + (\rho_{0,h} - y \rho_{0,h}) / c_{0, h}^{2},
            &
            \mathrm{if}\quad y \geq 0,
            \\
            \rho_{0,\ell} + (\rho_{0,h} - y \rho_{0,\ell}) / c_{0, \ell}^{2},
            &
            \mathrm{else},
        \end{cases}
        \\
        u(\mathbf{x}, t = 0)
        &=
        \begin{cases}
            \displaystyle+0.01\sin(k x) \frac{\cosh(k (y - 1.0))}{\sinh(k)}
            &
            \mathrm{if}\quad y \geq 0
            \\
            \displaystyle-0.01\sin(k x) \frac{\cosh(k (y + 1.0))}{\sinh(k)}
            &
            \mathrm{else},
        \end{cases}
        \\
        v(\mathbf{x}, t = 0)
        &=
        \begin{cases}
            \displaystyle-0.01\cos(k x) \frac{\sinh(k (y - 1.0))}{\sinh(k)}
            &
            \mathrm{if}\quad y \geq 0
            \\
            \displaystyle+0.01\cos(k x) \frac{\sinh(k (y + 1.0))}{\sinh(k)}
            &
            \mathrm{else},
        \end{cases}
    \end{align}
\end{subequations}
where $k = 4\pi$ is the perturbation wave number.
The linear growth rate of the RTI is, see, e.g.,~\cite{chandrasekhar2013hydrodynamic, choudhuri1998physics, wang2010combined},
\begin{equation}
    \omega_{\mathrm{RTI}} = \sqrt{g k \frac{\rho_{0, h} - \rho_{0, \ell}}{\rho_{0, h} + \rho_{0, \ell}}}.
\end{equation}
We normalize time using $T = 2\pi/\omega_{\mathrm{RTI}}$.
The reference density is $\rho_{0,h} = 1.0$ and $\rho_{0,\ell} = \Theta$ in the heavy and in the light phase, respectively.
The artificial speed of sound is $c_{0, h} = 14.1$ and $c_{0,\ell} = c_{0, h} \sqrt{1/\Theta}$.
We increase the speed of sound in the light phase to compensate for compressibility effects due to the decreasing density ratio, see, e.g.,~\cite{colagrossi2003numerical, hammani2020detailed}.
\par
The particle distribution for different density ratios is shown in Fig.~\ref{fig:RayleighTaylorParticles}.
We observe characteristic mushroom-shaped structures as the heavy phase accelerates into the light phase, and a transition to finger-like morphology at lower density ratios.
We find that, similar to the KHI, the particle distributions remain relatively isotropic even near the material interface and free of numerical artifacts, demonstrating the robustness of the multiphase GMFH scheme under gravitational stratification.
\begin{figure}[h!]
    \centering
    \includegraphics{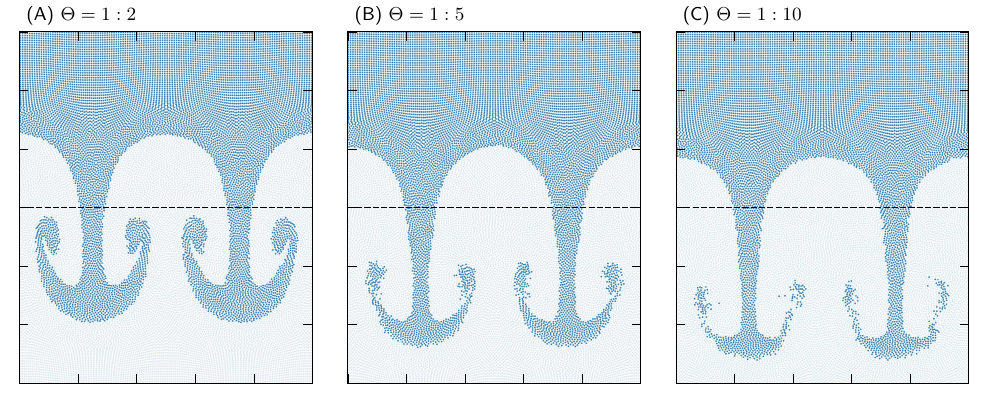}
    \caption{
        The particle distribution of the light phase ($\textcolor{Blues3-1}{\bullet}$) and heavy phase ($\textcolor{Blues3-3}{\bullet}$) for the RTI with density ratio (\textsf{A}) $\Theta = 1:2$, (\textsf{B}) $\Theta = 1:5$, and (\textsf{C}) $\Theta = 1:10$ at particle resolution $\Delta x_{0} = 1 / 128$ and normalized time $\tau = 1.2$.
        The black dashed line (--\,--) represents the initial position of the material interface.
    }
    \label{fig:RayleighTaylorParticles}
\end{figure}
\par
The mode amplitude for different density ratios and particle resolution is shown in Fig.~\ref{fig:RayleighTaylorMode}.
We compute the mode amplitude for the investigated wave number using discrete convolution (as done for the KHI by McNally et al.~\cite{mcnally2012wellposed}) and compare it with the exact relative amplitude in the linear growth regime.
The accuracy of the numerical solution improves monotonically with resolution in the linear growth regime, though at a markedly slower rate than for the KHI.
\begin{figure}[h!]
    \centering
    \includegraphics{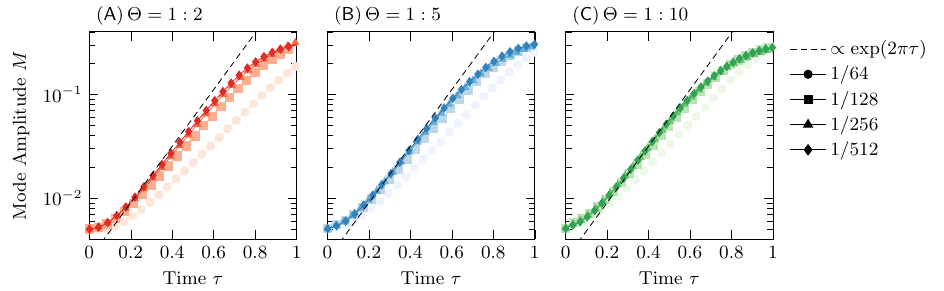}
    \caption{
        The linear mode amplitude $M$ of the $k = 4 \pi$ perturbation mode over non-dimensional time $\tau$ for the RTI for $\mathsf{(A)}$ $\Theta = 1 : 2$, $\mathsf{(B)}$ $\Theta = 1 : 5$ and $\mathsf{(C)}$ $\Theta = 1 : 10$ at particle resolution $\Delta x_{0} = 1/64$ (\raisebox{-0.2ex}{\scalebox{1.3}{$\bullet$}}), $\Delta x_{0} = 1/128$ (\scalebox{0.8}{$\blacksquare$}), $\Delta x_{0} = 1/256$ ($\blacktriangle$), and $\Delta x_{0} = 1/512$ ($\blacklozenge$).
        The black dashed line (--\,--) represents the exact relative amplitude in the linear growth regime.
    }
    \label{fig:RayleighTaylorMode}
\end{figure}
\par
The smoothed color function for different density ratios at the finest particle resolution is shown in Fig.~\ref{fig:RayleighTaylorColor}.
The proposed multiphase GMFH scheme resolves large-scale RTI structures and secondary small-scale hydrodynamic instabilities with high fidelity.
The RTI exhibits symmetry-breaking in the limit of vanishing numerical viscosity with increased resolution, similar to grid-based schemes, see, e.g.,~\cite{fu2019very, fu2021very}.
A fully symmetry-preserving algorithm, similar to Fleischmann et al.~\cite{fleischmann2019numerical}, is not feasible for GMFH schemes.
\begin{figure}[h!]
    \centering
    \includegraphics{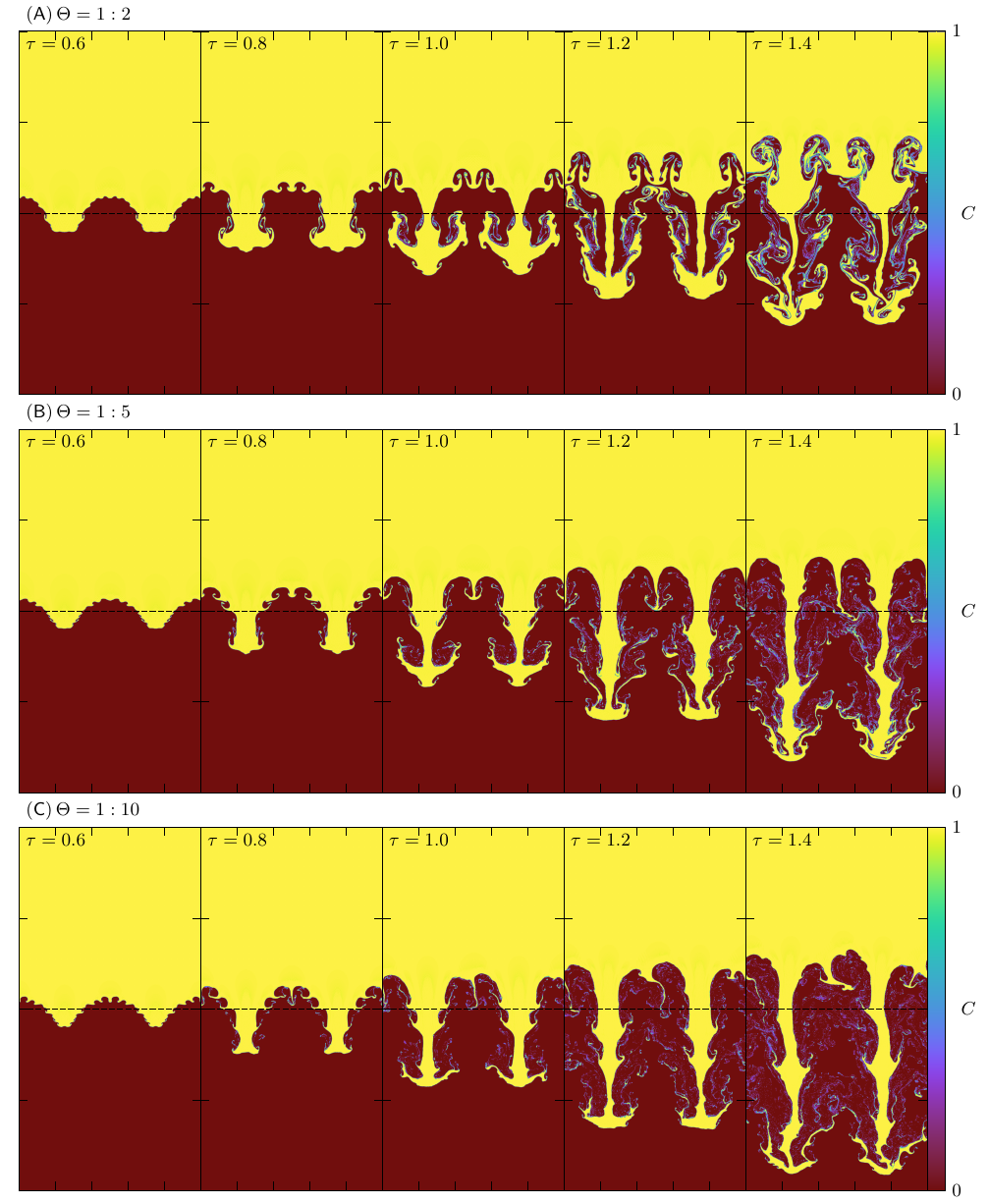}
    \caption{
        The smoothed color function $C$ for the RTI for the density ratios $\mathsf{(A)}$ $\Theta = 1 : 2$, $\mathsf{(B)}$ $\Theta = 1 : 5$ and $\mathsf{(C)}$ $\Theta = 1 : 10$ at different non-dimensional time for particle resolution $\Delta x_{0} = 1/1024$.
        The black dashed line (--\,--) represents the initial position of the material interface.
    }
    \label{fig:RayleighTaylorColor}
\end{figure}
\section{Conclusion \& Outlook}\label{sec:Conclusion}
In this work, we have developed a GMFH scheme for weakly compressible multiphase flows.
The framework is built upon the MFM/MFV scheme of Hopkins~\cite{hopkins2015new} and generalizes the multiphase SPH scheme of Hu and Adams~\cite{hu2006multi}.
The method employs a second-order accurate reconstruction of the primitive variables onto the moving particle interface and introduces sufficient upwinding through the solution of the rotated Riemann problem using a subsonic HLLC Riemann solver.
Following Gaburov and Nitadori~\cite{gaburov2011astrophysical}, we use the renormalized gradient formulation of Lanson and Vila~\cite{lanson2008renormalized, lanson2008renormalized2} to increase accuracy.
We advect the particles using the quasi-Lagrangian material velocity of Michel et al.~\cite{michel2022particle}, which preserves the Lagrangian fluid velocity in an infinite resolution limit and introduces a relaxation mechanism that improves the particle distribution for increased accuracy and stability at finite particle resolution.
We treat the material interface consistently to suppress mass fluxes over the material interface.
\par
We validated our methodology using several challenging benchmark problems.
The Taylor-Green vortex shows that total mass and linear momentum are conserved to machine precision.
Angular momentum conservation and kinetic energy dissipation improve with increasing resolution.
The material interface advection problem confirms that the scheme preserves hydrodynamic stability for extreme density ratios without artificial mixing or spurious particle penetration.
The surface gravity wave problem demonstrates the method's ability to capture interface dynamics under propagating surface gravity waves near hydrostatic equilibrium.
Finally, the Kelvin-Helmholtz and Rayleigh-Taylor instabilities show that the scheme correctly predicts unstable wave growth.
\par
The proposed GMFH scheme has a higher computational cost than standard SPH due to explicit computations, renormalization, and limiting of the primitive gradients, and the need to solve a rotated Riemann problem.
Several simplifications trade accuracy for reduced cost.
As shown above, the method provides a generalized SPH formulation, see Eq.~\eqref{eq:EffectiveInterParticleSurface}, eliminating the need for gradient renormalization.
Replacing the second-order reconstruction with a first-order scheme further reduces complexity.
Adopting the low-dissipation Riemann solver of Zhang et al.~\cite{zhang2017weakly} reduces the excessive numerical dissipation that would otherwise occur.
If computational cost is a limiting factor in practical applications, these modifications narrow the gap toward classical SPH at the expense of formal accuracy.
\par
The proposed method requires a fully supported particle neighborhood, similar to the approach of Hu and Adams~\cite{hu2006multi}, and therefore does not apply directly to free-surface flows.
However, any free-surface problem can be reformulated as a two-phase problem by explicitly retaining both phases, at the expense of an increased particle count and computational cost.
Furthermore, free-surface models are often adopted precisely because a robust multiphase treatment of very low density ratios has not been available.
We have demonstrated robustness at extreme density ratios, showing that the present method bridges this gap.
\par
Several natural extensions of the present framework are possible.
Surface tension could be incorporated using the continuum surface force (CSF) approach of Brackbill~\cite{brackbill1992continuum}, as previously adapted to SPH~\cite{morris2000simulating,hu2006multi,adami2010new,zoeller2023partitioned}.
Because the material interface naturally remains at the finest resolution under Lagrangian advection, the framework is also well suited to a multiresolution scheme that concentrates resolution at the interface without dynamic refinement.
Finally, extending the formalism to fully compressible multiphase flows, including a hybrid formulation coupling a fully compressible light phase to a weakly compressible heavy phase, would open the method to violent shock-interface interactions relevant to, e.g., droplet aerobreakup and jet atomization.
\section*{Acknowledgements}
The authors would like to thank the TUM-Oerlikon Advanced Manufacturing Institute (AMI) for funding this project.
\section*{CRediT authorship contribution statement}
\textbf{Fabian Fritz:} Writing – original draft, Visualization, Validation, Software, Methodology; \textbf{Nikolaus A. Adams:} Writing – review \& editing, Supervision, Funding acquisition; \textbf{Stefan Adami:} Writing – review \& editing, Supervision, Methodology.
\section*{Declaration of competing interest}
The authors declare that they have no known competing financial interests or personal relationships that could have appeared to
influence the work reported in this paper.
\section*{Data availability}
No data was used for the research described in the article.
\bibliographystyle{plain}
\bibliography{references}
\appendix
\section{A comparison between single and double precision accuracy}\label{sec:FloatingPointAccuracy}
Single precision (FP32) and double precision (FP64) accuracy refer to two different formats for storing floating-point values in modern computing architectures~\cite{goldberg1991every}.%
They differ in terms of accuracy and memory usage: %
FP32 uses 32 bits and provides approximately 7 decimal digits of accuracy, making it faster and more memory-efficient, but less precise. %
FP64 uses 64 bits and offers around 15 to 17 decimal digits of accuracy. %
This improved accuracy enhances numerical stability and allows for a wider dynamic range, but it comes at the cost of higher computational and memory demands. %
Formally, FP64 performance is about half of the FP32 performance. %
The rise of GPUs in scientific computing has reignited an interest in FP32 accuracy, as most consumer GPUs do not support FP64 computations natively or have very few FP64 execution units. %
This drastically reduces the performance between FP32 and FP64. %
We report 1/16 of the performance on an NVIDIA RTX A6000 and a performance as low as 1/64 on an NVIDIA RTX 2080Ti when using FP64 instead of FP32 for our specific implementation. %
This performance penalty on many GPUs has motivated researchers to exploit (and possibly abuse) FP32 precision in scientific computing, see, e.g., Zhao et al.~\cite{zhao2023multi} -- without implying that their chosen precision was insufficient. %
However, other researchers do not state the chosen precision at all. %
In the following section, we compare the accuracy of FP32 and FP64 for the proposed GMFH scheme. %
The goal is to evaluate whether FP32 precision is sufficient and, if so, under which conditions. %
We do not consider a mixed-precision approach in this work. %
\par
A comparison between the particle distribution is given in Fig.~\ref{fig:RTI2dPrecision} for two different resolutions of the RTI. %
In the following, we summarize our findings: 
\begin{itemize}
    \item %
    The particle distribution between FP32 and FP64 accuracy does not differ significantly for the moderate resolution $\Delta x_{0} = 1 / 256$. %
    Thus, for this resolution and coarser resolutions, FP32 accuracy is sufficient. %
    \item %
    The particle distribution between the FP32 and FP64 representation differs significantly for the high resolution $\Delta x_{0} = 1 / 512$.
    The FP64 representation looks like a converged particle distribution, while the FP32 representation shows strong symmetry break.
    Thus, for this resolution and finer resolution, FP32 accuracy is no longer sufficient.
\end{itemize}
We can make no general recommendation on whether FP32 accuracy yields sufficient precision, as it depends on several factors, such as, e.g., resolution, fluid states, material parameters, and domain size.
The impact of floating-point inaccuracies depends heavily on specific implementation details and, while it can be mitigated, cannot be entirely eliminated.
We need to determine the precision in a case-by-case fashion.
Overall, we discourage the use of FP32 accuracy despite the performance penalty associated with FP64 accuracy.
\begin{figure}[h]
    \centering
    \includegraphics[page=1]{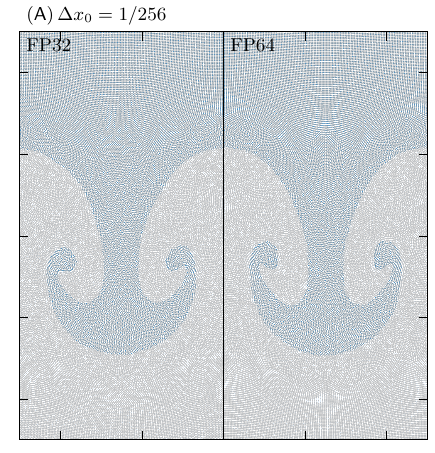}
    \includegraphics[page=2]{figures/rayleigh-taylor-instability_2d_precision.pdf}
    \caption{
        The particle distribution of the light phase ($\textcolor{Blues3-1}{\bullet}$) and heavy phase ($\textcolor{Blues3-3}{\bullet}$) for the RTI using a quasi-Lagrangian scheme for density ratio $\Theta = 1:2$ at $\tau = 1.0$ using (\textsf{A}) $\Delta x_{0} = 1 / 256$ and (\textsf{B}) $\Delta x_{0} = 1 / 512$.
        The left plane shows the single precision (FP32) results and the right plane the double precision (FP64) results.
    }
    \label{fig:RTI2dPrecision}
\end{figure}

\end{document}